\documentclass[%
reprint,
amsmath,amssymb,
aps,
pra,
]{revtex4-2}
\usepackage{graphicx}
\usepackage{dcolumn}
\usepackage{bm}
\usepackage[colorlinks=true,linkcolor=red,citecolor=blue]{hyperref}
\usepackage[mathlines]{lineno}

\usepackage{bbm}
\usepackage{dsfont}
\usepackage{amsmath}
\usepackage{verbatim}
\usepackage{graphicx}
\usepackage{braket,amsmath}
\usepackage[dvipsnames, svgnames, x11names]{xcolor}
\usepackage{array}
\usepackage{bibunits}
\usepackage{braket}
\usepackage{mathrsfs}
\usepackage{amsmath}
\usepackage{graphicx}
\usepackage{dcolumn}
\usepackage{bm}
\usepackage[capitalise]{cleveref}
\crefname{section}{Sec.}{Secs.}
\crefname{appendix}{App.}{Apps.}

\begin{document}
	
	\preprint{APS/123-QED}
	
	\title{Quantum teleportation between a continuous-variable optical qumode and a discrete-variable solid-state qubit
}
	

		\author{Di Wang}%
	\affiliation{%
		Department of Physics, Wenzhou University, Zhejiang 325035, China
	}%

\author{Lei Xie}%
	\affiliation{%
		Department of Physics, Wenzhou University, Zhejiang 325035, China
	}%

	\author{Jinfeng Liu}%
	\affiliation{%
		Department of Physics, Wenzhou University, Zhejiang 325035, China
	}%
		\author{Yiling Song}%
	\affiliation{%
		Department of Physics, Wenzhou University, Zhejiang 325035, China
	}%
		
	\author{Wei Xiong}
	\affiliation{%
		Department of Physics, Wenzhou University, Zhejiang 325035, China
	}%
	\author{Mingfeng Wang}
	\email{mfwang@wzu.edu.cn}
	\affiliation{%
		Department of Physics, Wenzhou University, Zhejiang 325035, China
	}%
	
	
	\date{\today}
	
	\begin{abstract}
Quantum teleportation is a fundamental ingredient for quantum information science and technology. In particular, the ability to perform quantum teleportation between quantum systems of different natures and encoding types is crucial for building complex systems, such as distributed quantum internet. Here we propose a scheme to teleport a continuous variable optical qubit, encoded in an optical qumode by means of a superposed coherent state, onto a discrete variable solid-state qubit, associated with a single nitrogen-vacancy center spin in diamond, via a hybrid entanglement. By using a newly developed method for Bell-state measurement, which relies only on light homodyne detection and spin polarization measurement, near-deterministic and -perfect quantum teleportation can be achieved for large coherent-state amplitude input. Taking noise effects into account, we find that the average teleportation fidelity can still exceed the classical limit, enabling substantial teleportation distances under realistic experimental conditions.
		\begin{description}
			\item[PACS numbers]
			 03.67.-a, 03.65.Ta, 42.50.Pq
		\end{description}
	\end{abstract}
	
	
	\pacs{Valid PACS appear here}
	\maketitle
	
	
	\section{\label{sec:level1}INTRODUCTION}
   In realistic applications of quantum technologies \cite{nielsen2001quantum}, different natures of physical systems are usually required. For example, solid-state qubits, such as superconducting qubits \cite{devoret2013superconducting} and nitrogen-vacancy (NV) centers \cite{nemoto2014photonic}, are used to perform quantum gate operations in quantum computing or used for quantum information storage, while photons \cite{gisin2007quantum} serve as transmission media to connect different quantum qubits, enabling the remote quantum communication between quantum network nodes \cite{reiserer2022colloquium}. In order to achieve systematic quantum information processing (QIP), it is essential to integrate various physical systems in a hybrid fashion \cite{andersen2015hybrid}.

   In the field of QIP, quantum information can be encoded either as discrete variables (DV), such as the two-level system known as qubit \cite{bouwmeester1997experimental}, or as continuous variables (CV), like the quadrature amplitudes of an optical mode, known as qumode \cite{braunstein1998teleportation}. For DV qubits, it is relatively easy to prepare and manipulate quantum states, as well as to implement quantum logic gates, making them suitable for precise quantum computing \cite{kok2007linear}. While CV encoding possess the advantages of efficient compatibility and scalability, which makes it suitable for the construction of large-scale quantum networks and the realization of high-performance quantum communication \cite{braunstein2005quantum}. Recently, it has been shown that the combination of these two types of encoding can offer protocols that outperform both pure CV and DV solutions \cite{andersen2015hybrid}. More specifically, using hybrid methods in quantum computing \cite{omkar2020resource,gan2020hybrid} and quantum networks \cite{sheng2013hybrid,pirandola2016physics} bears favourable results. For example, Lee et al. \cite{lee2013near} proposed a method using all-optical hybrid encoding to achieve universal quantum computation, which is superior to the linear optical
   quantum computing \cite{knill2001scheme,nemoto2004nearly} and the coherent-state
   quantum computing \cite{jeong2001quantum,ralph2003quantum,lund2008fault}. Takeda et al. \cite{takeda2013deterministic} demonstrated the CV quantum teleportation of DV states, showing that such hybrid method possesses the great advantage of being deterministic and solely linear-components dependence.
   Therefore, the development of hybrid techniques and the ability to convert quantum information between different encoding platforms is crucial for building complex systems, such as quantum internet.

   Currently, a great effort has been oriented to develop methods for transferring quantum state from DV to CV qubits \cite{jeong2014generation,morin2014remote,huang2019engineering,ulanov2015undoing,gao2013quantum}.  Park et al. \cite{park2012quantum} proposed to transfer optical state between polarization and coherent-state qubits via quantum teleportation. Most recently, Ulanov et al. \cite{ulanov2017quantum} demonstrated a protocol that utilises hybrid entanglement resources as quantum channel to teleport a CV qubit, encoded as a superposition of optical coherent states, onto a DV qubit that is a superposition of the vacuum and single-photon states.
   These protocols, however, have primarily focused on quantum-state conversion between CV and DV platform within a single type of quantum system---specifically, light. Indeed, efficiently transferring quantum states over \emph{long distances} between physical systems of \emph{different nature} is even more crucial for practical quantum communication and quantum networks. In particular, the quantum state transfer from a CV optical oscillator to a DV spin system plays a key role in QIP, since such process corresponds to the conversion of flying photon qubits to stationary spin qubits--- an important way to realize long-time quantum memory for quantum information encoded in light \cite{gao2013quantum}.

   In this paper, we propose a scheme to teleport a CV optical qubit, encoded in an optical qumode by means of a superposed coherent state, onto a DV spin qubit, encoded in two levels of an NV center in diamond, via a hybrid entangled resource. The NV center in diamond has sparked widespread research interest due to its ability to maintain both the purity and a long spin lifetime, as well as its controllability at room temperature \cite{doherty2013nitrogen,buckley2010spin,balasubramanian2009ultralong}, which has been extensively used as memory units to store quantum information for relatively long periods \cite{nakazato2022quantum,abobeih2018one,childress2006coherent}. Our proposed scheme relies on the hybrid entanglement between an NV center (DV system) and a light pulse (CV system), which can be created deterministically by simply reflecting a coherent laser pulse from an optical cavity containing a single NV center. Besides the quantum channel, the Bell-state measurement (BSM) is also an important ingredient for the quantum-teleportation protocol. We developed a new approach to achieve near-deterministic BSM. By first combining the sender's part of quantum channel and the state to be teleported on a balanced beam splitter, and then sending the output modes to interact with two auxiliary NV centers, we can achieve BSM solely through homodyne detections (HD) of light modes and measurements of the spin direction. Compared with the previous approaches, neither photon-number parity measurement \cite{lee2013near} nor nonlinear Kerr interaction \cite{liao2006new} is required, which greatly simplifies the realistic implementation. By utilizing the newly developed BSM, we demonstrate that it is possible to achieve nearly deterministic and nearly perfect quantum teleportation from a flying qumode to a stationary qubit under certain conditions.

   The rest of the paper is organized as follows. In \cref{sec:level2} we will first introduce the spin system and then give the principle of spin-light interaction. In \cref{sec:level2.2} we present the details of the proposed protocol. In \cref{sec:level2.3} we will consider the influence of noise effects. Finally, we summarize in \cref{sec:level3}.
 \section{\label{sec:level2}The model}
\begin{figure}[t]
    	\includegraphics[scale=0.40]{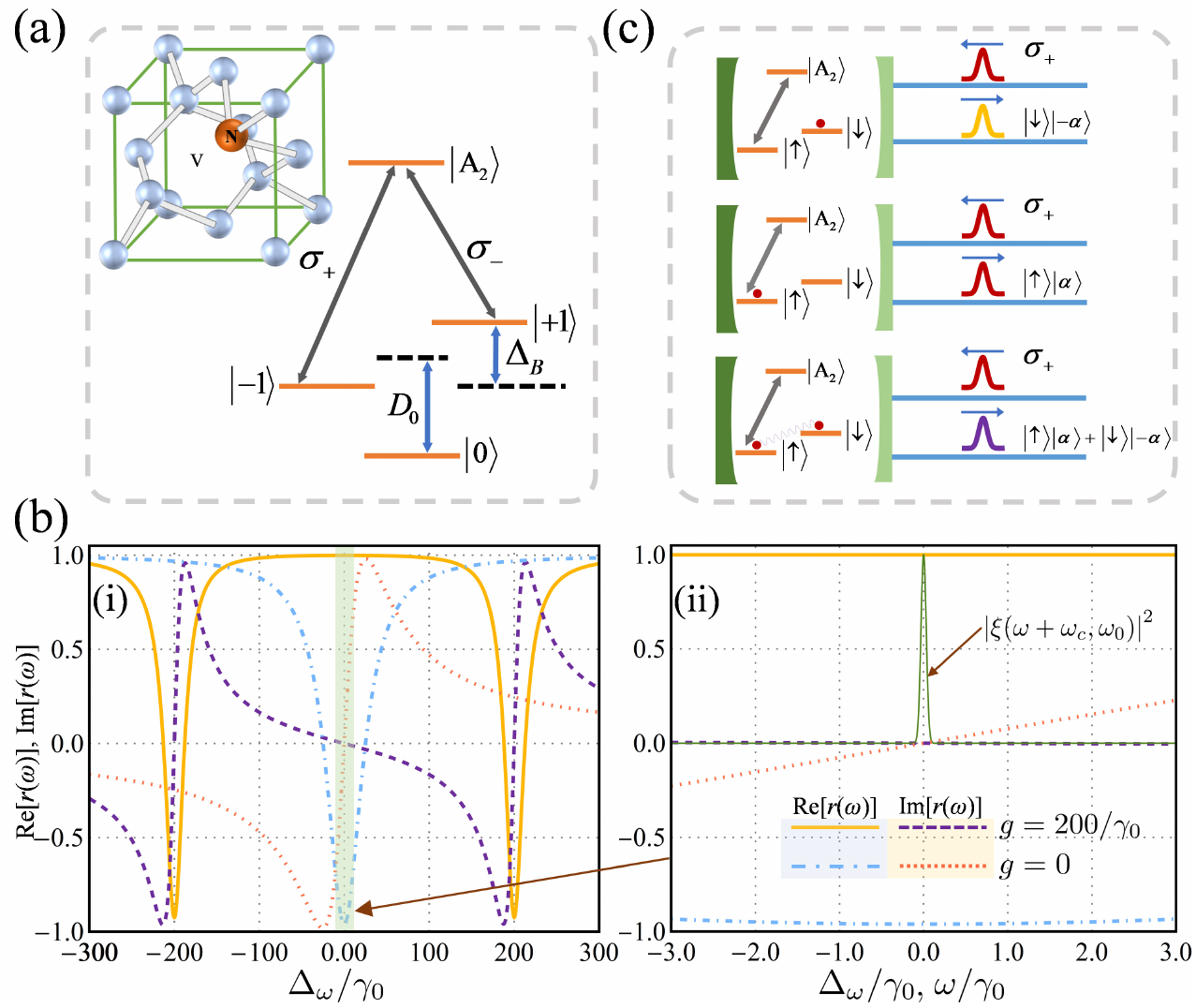}
    	\caption{(a) Energy level diagram of the NV center. It has three ground-state sublevels $\ket{0},\ket{\pm1}$, and we choose $\ket{A_2}$ as the excited state, which couples to the ground states $\ket{-1}$ and $\ket{+1}$ via the left polarized light $\sigma_+$ and the right polarized light $\sigma_-$, respectively. (b) The reflection coefficient $r_\omega$ as a function of detuning $\Delta_\omega$. The frequency of the input light pulse is narrow bandwidth and near resonant with the cavity ($\omega_0\simeq\omega_c$), then the imaginary part of $r_\omega$ can be neglected and every frequency mode $\hat a(\omega)$ of the input photon packet will experience a $\pi$-phase shift depending on whether the cavity mode interacts with the NV center or not. (c) The CP gate operations for NV spin and input photonic pulse. If the spin is initially prepared in $\ket{\downarrow}$, the photon packet is reflected with a $\pi$-phase shift. If the spin is in $\ket{\uparrow}$, the photon packet is reflected with no phase shift. Finally, if the ground state is prepared in the superposition state $\ket{\psi_S}$, the reflected photons get entangled with spin.}\label{fig1}
    \end{figure}

   We consider a cavity system containing an NV center, which couples to an optical coherent pulse. The ground state ${^3\kern-0.3em}A_2$ of the negatively charged NV center is a triplet with spin $S=1$, i.e., $\left|{m_s} = {0, \pm 1} \right\rangle$, as shown in \cref{fig1}(a). The (zero-field) splitting between the sublevels $\left| 0 \right\rangle$ and $\left| { \pm 1} \right\rangle$ is about ${D_0} = 2.87$ GHz \cite{teissier2014strain,marcos2010coupling}. In the presence of a magnetic field, the degeneracy between the $\left| +1 \right\rangle$ and $\left| -1 \right\rangle$ can be lifted by an amount ${\Delta _B} = {g_e}{\mu _B}{B_z}/\hbar$ \cite{bennett2013phonon}. The excited state ${^3\kern-0.2em}E$ of an NV center is also a spin triplet \cite{yang2010entanglement}. Due to the spin-orbit interaction, the spin-spin interactions, and the nonaxial strain, the degeneracy of ${^3\kern-0.2em}E$ will be lifted to yield six electronic excited states \cite{chen2011entangling,wei2015hybrid}. Among them, the excited state
   \begin{eqnarray}
   	\left| {{A}_{2}} \right\rangle =\frac{1}{\sqrt{2}}(\left| {{E}_{-}} \right\rangle \left| +1 \right\rangle +\left| {{E}_{+}} \right\rangle \left| -1 \right\rangle ),\label{eq1}
   \end{eqnarray}
    which is an entangled state of spin states and orbital states $\ket{E_\pm}$, is robust with the symmetric properties \cite{ren2017robust}. Due to total angular momentum conservation, the $\sigma^+$ photon can only drive the transition $\ket{-1}\rightarrow\ket{A_2}$, while the $\sigma^-$ photon can induce the transition $\ket{+1}\rightarrow\ket{A_2}$ \cite{li2020heralded}. In this study, the qubit will be encoded in the sublevels $\{\ket{\uparrow} \equiv\ket{-1},\ket{\downarrow} \equiv\ket{+1}\}$ while the excited state $\ket{A_2}$ plays the role of ancillary.

    As illustrated in \cref{fig1}(c), our protocol relies on the interaction between an optical pulse and a cavity system containing an NV center described above. We assume that the spin system is initially prepared in the ground subspace $\{\ket{\uparrow},\ket{\downarrow}\}$, which can be realized by first optical pumping to the ground sublevel $\ket{0}$ and then applying a $\pi$ pulse to achieve the transition $\ket{0}\rightarrow \ket{\uparrow}$ \cite{neumann2010single}. Next, a $\pi/2$ pulse driven the transition $\ket{\uparrow}\rightarrow\ket{\downarrow}$ is applied to generate an equal superposition state $\ket{\psi_S}=(\ket{\downarrow}+\ket{\uparrow})/\sqrt{2}$. The spin transition $\ket{\uparrow}\rightarrow\ket{A_2}$ is coupled to a left-circularly polarized cavity mode $\hat a$ with frequency $\omega_c$, which is driven by a left-circularly polarized pulse that is in a coherent state $\left| \alpha  \right\rangle  = \exp ( - {\left| \alpha  \right|^2}/2)\exp [\alpha \int {d\omega \xi (\omega ,{\omega _0})\hat a_{\omega}^\dag } ]\left| 0 \right\rangle $ \cite{duan2005robust},  where $\xi(\omega,\omega_0)$ is the normalized pulse amplitude spectrum with $\omega_0$ being the central frequency of the spectrum, $\hat a_{\omega}$ is a one-dimensional field, satisfying the commutation relation $[{{\hat a}_{\omega}} ,\hat a_{\omega'}^{\dagger } ]=\delta (\omega-\omega')$, and $\ket{0}$ stands for the vacuum state of all the optical modes  \cite{walls1994quantum}. The average photon number of the
    pulse is given by ${{\left| \alpha  \right|}^{2}}$. The dynamics of the intracavity can be described by the Hamiltonian
       \begin{figure*}[t]
\includegraphics[scale=0.6]{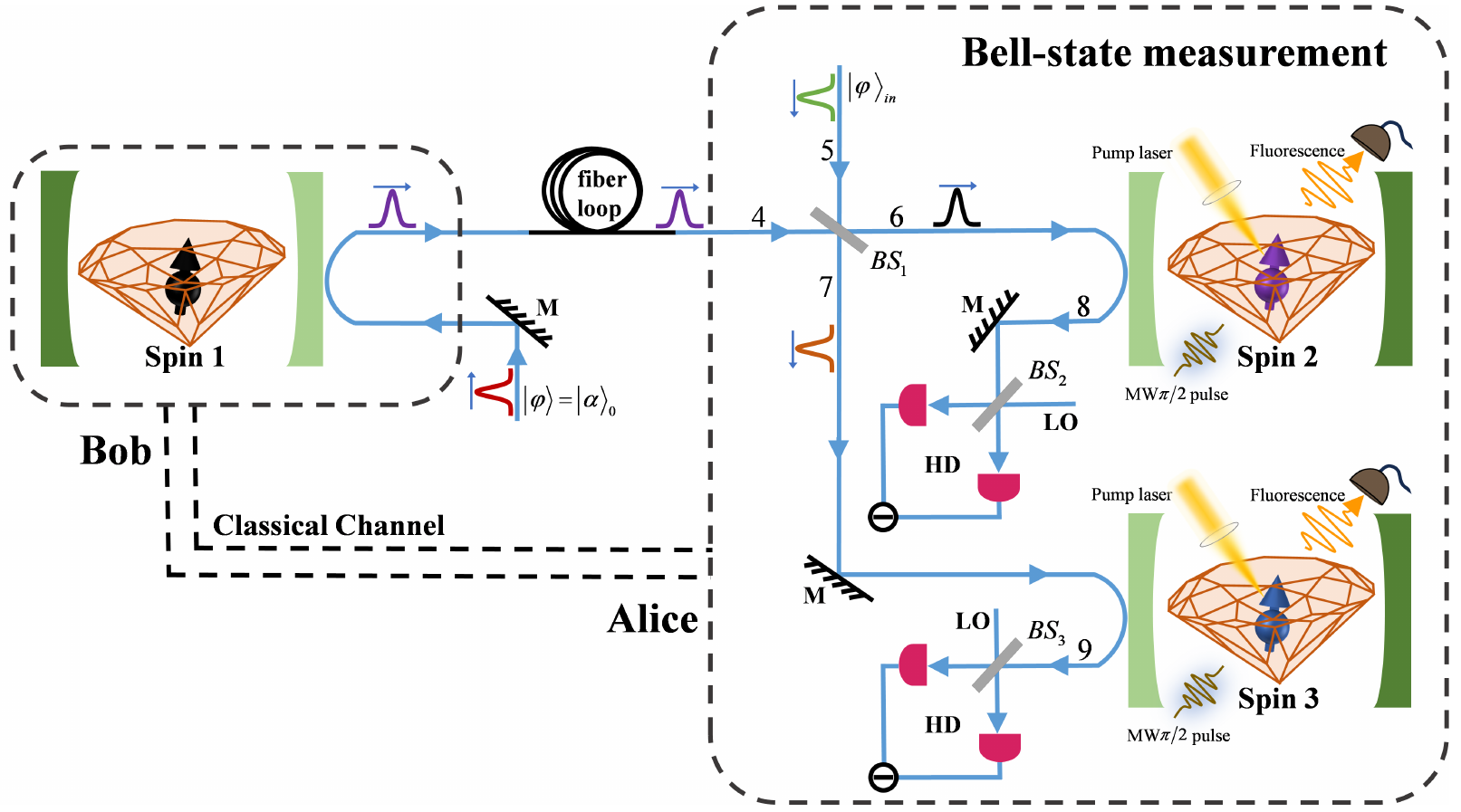}
\caption{An unknown quantum state $\ket{\varphi}_{\rm{in}}$ carried by an optical pulse at Alice's site will be teleported onto the NV spin $1$ contained in an optical cavity at Bob's site. To do so, Bob sends an optical pulse in state $\ket{\varphi}$ to interact with the spin-cavity system. After the reflection by the cavity wall, the optical pulse gets entangled with spin 1 [which can be described by Eq. (\ref{eq11})], forming the quantum channel. The optical pulse is then directed into a fiber, and after passing through the fiber, it will be combined with the information pulse at $B{S_1}$. Then, the output pulses $6$ and $7$ of the two ports of $B{S_1}$ are sent to interact with two auxiliary NV spins 2 and 3, respectively. Finally, the BSM is performed through a measurement of the spin polarization (via spin-to-charge conversion \cite{giri2023charge}) and a HD detection of the reflected light modes. Alice will send her measurement results to Bob through a classical channel. According to the information received, a unitary transformation $\hat U_o$ (shown in Table \ref{table1}) will be applied to the spin 1 to complete the teleportation. BS: beam splitter; LO: local oscillator; MW: microwave.}\label{fig2}
\end{figure*}
    ($\hbar=1$)
       \begin{eqnarray}
    	\hat H=\frac{{\omega_a}}{2}{{\hat\sigma }_{z}}+{{\omega }_{\text{c}}}{{\hat a}^{\dagger }}\hat a+g\left( {{\hat a}^{\dagger }}{{\hat{\sigma} }_{-}}\text{+}\hat a{{\hat{\sigma} }_{+}} \right),\label{eq2}
    \end{eqnarray}
    where $\hat{\sigma}_z=\ket{A_2}\bra{A_2}-\ket{\uparrow}\bra{\uparrow}$ and $\hat{\sigma}{_+}=\ket{A_2}\bra{\uparrow}$  ($\hat{\sigma}{_-}=\hat{\sigma}_+^\dag$) are the Pauli matrix operators, and $\omega_a$ refers to the frequency difference between $\ket{\uparrow}$ and $\ket{A_2}$, and $g$ denotes the coupling constant. Next, we assume that the spin-cavity system is driven by an optical field with frequency $\omega$. In the rotating frame with respect to $\omega$ the Heisenberg equations of motion for the cavity mode and atomic lowering operator are given by \cite{hu2008giant}
        \begin{eqnarray}
    	{\dot{\hat a}\left( t \right)} &=&  - \left[ {i\left(\omega_c-\omega\right)  + \frac{\kappa }{2} + \frac{\eta }{2}} \right]\hat a\left( t \right)\nonumber\\
    	&&- g{{\hat \sigma }_ - }\left( t \right) - \sqrt \kappa  {{\hat a}_{{\rm{in}}}}\left( t \right),\label{eq3}\\
    	{{\dot{\hat\sigma} }_ - }\left( t \right) &=&  - \left[ {i\left(\omega_a-\omega\right)  + \frac{\gamma_0 }{2}} \right]{{\hat \sigma }_ - }\left( t \right)\nonumber\\
    	&&- g{{\hat \sigma }_z}\left( t \right)\hat a\left( t \right) + \sqrt \gamma_0  {{\hat \sigma }_z}\left( t \right){{\hat b}_{{\rm{in}}}}\left( t \right),\label{eq4}
    \end{eqnarray}
    where $\kappa$ is the cavity decay rate, and $\eta$ is the decay rate of light to bath, and $\gamma_0$ is the spin decay rate. $\hat a_{\rm{in}}$ and $\hat b_{\rm{in}}$ denote the input field operator and the vacuum noise operator, respectively, satisfying the commutation relations $[ {{\hat a}_{\rm{in}}}(t), \hat a_{\rm{in}}^{\dagger }({t}') ]=[ {{\hat b}_{\rm{in}}}(t), \hat b_{\rm{in}}^{\dagger }({t}') ]=\delta (t-{t}')$. In the limit of weak excitation ($\left\langle {{\hat\sigma _z}} \right\rangle  \simeq  - 1$), one may adiabatically eliminate the cavity mode and obtain the reflection coefficient $r_\omega=\frac{\hat a_{\rm{out}}(t)}{\hat a_{\rm{in}}(t)}$ for the spin-cavity system \cite{chen2011entangling,an2009quantum}
          \begin{eqnarray}
    r_\omega = \frac{{4\left( {{g^2} - \Delta _\omega ^2} \right) + \gamma_0 \left( {\eta  - \kappa } \right) + i2{\Delta _\omega }\left( {\gamma_0  + \eta  - \kappa } \right)}}{{4\left( {{g^2} - \Delta _\omega ^2} \right) + \gamma_0 \left( {\eta  + \kappa } \right) + i2{\Delta _\omega }\left( {\gamma_0  + \eta  + \kappa } \right)}}\label{eq5}
\end{eqnarray}
   with $\Delta_\omega=\omega_c-\omega$, where we have used the standard input-output relation $	{{\hat{a}}_{\rm{out}}}(t)={{\hat{a}}_{\rm{in}}}(t)+\sqrt{\kappa }\hat{a}(t)$ and assumed that the cavity mode is on resonance with $\ket{\uparrow}\rightarrow\ket{A_2}$ transition ($\omega_c=\omega_a$). Figure \ref{fig1}(b)(i) plots the reflection coefficient as a function of detuning $\Delta_\omega$, showing that in the regime of small detuning ($\Delta_\omega\simeq 0$) $r_\omega$ is mainly determined by its real part. If the cavity mode interacts strongly with the NV center ($g\gg \{\kappa,\gamma_0,\eta\}$), then one has $r_\omega\simeq 1$ and thus $\hat a_{\rm{out}}(t)\simeq \hat a_{\rm{in}}(t)$, indicating that the light mode is reflected without phase change. On the contrary, if the cavity mode dose not interact with the NV center ($g=0$), we have $\hat a_{\rm{out}}(t)\simeq -\hat a_{\rm{in}}(t)$, showing that the reflected light mode experiences a $\pi$-phase shift.

   The input optical pulse involved here is assumed to be narrow bandwidth and near resonant with the cavity, as shown in Fig. \ref{fig1}(b)(ii). Then if the spin is initially prepared in either $\ket{\downarrow}$ (corresponding to $g=0$) or $\ket{\uparrow}$ (corresponding to large $g$), the output state would be a separable state: $\ket{\downarrow}\ket{-\alpha}$ or $\ket{\uparrow}\ket{\alpha}$ [see Fig. \ref{fig1}(c)], respectively. However, if the initial state is in the superposition state $\ket{\psi_S}$, then the reflection process drives the system into the entangled state
      $\frac{1}{\sqrt{2}}\left(\left| \uparrow  \right\rangle {{\left|  \alpha \right\rangle }}+\left| \downarrow  \right\rangle {{\left| -\alpha  \right\rangle }}\right)$, which, as we will show later, is the core ingredient for current scheme. Thus, one can conclude that the reflection of the optical pulse by the previously described spin-cavity system is equivalent to implementing a controlled phase (CP) transformation between the NV spin, say $i$, and the optical mode, say $j$, which can be described by the unitary operator $\hat U^{CP}_{ij}=e^{i\pi\ket{\downarrow}_{\kern-0.1em ii \kern-0.1em}\bra{\downarrow}\otimes\int {d\omega \hat a_{\omega,j}^\dag } {\hat a_{\omega,j} }}$. Given the input states $\ket{\uparrow}_i\ket{\alpha}_j$ and $\ket{\downarrow}_i\ket{\alpha}_j$, the states of the hybrid
system after the applications of the CP gate are
\begin{eqnarray}
{{\hat U}^{CP}_{ij}}\left(\left|  \uparrow \right\rangle_i\otimes \left| \alpha  \right\rangle_j  \right)&=& \left|  \uparrow  \right\rangle_i \left| \alpha  \right\rangle_j ,\label{eq6}\\
{{\hat U}^{CP}_{ij}}\left(\left|  \downarrow  \right\rangle_i \otimes\left| \alpha  \right\rangle_j \right) &=& \left|  \downarrow  \right\rangle_i \left| { - \alpha } \right\rangle_j .\label{eq7}
\end{eqnarray}
These equations form the basis for describing the proposed protocol.

\section{\label{sec:level2.2} The protocol}	
   Next, we begin to show our teleportation protocol. The quantum state to be teleported is an arbitrary superposition of two equal-amplitude and anti-phase coherent states
   \begin{eqnarray}
		{{\left|\varphi\right\rangle }_{\rm{in}}}=\mathcal{N}\left(a\left|\beta\right\rangle +b\left|-\beta\right\rangle\right) ,\label{eq8}
	\end{eqnarray}
    where the norm $\mathcal{N}=[1+2 e^{-2|\beta|^2} \operatorname{Re}(a b^*)]^{-1 / 2}$, and the adjustable weight parameters satisfy $|a|^2+|b|^2=1$, and the coherent state $\ket{\beta}$ in the Fock-state basis is \cite{liao2006new}
     		\begin{eqnarray}
			\left| \beta  \right\rangle \text{=}{{e}^{{\scriptstyle{}^{-{{\left| \beta  \right|}^{2}}}/{}_{2}}}}\sum\limits_{n=0}^{\infty }{\frac{{{\beta }^{n}}}{\sqrt{n!}}}\left| n \right\rangle,\label{eq9}
		\end{eqnarray}
     and in the position representation is \cite{leonhardt1997measuring}
       \begin{eqnarray}
\left| \beta  \right\rangle  = \frac{1}{{{\pi ^{1/4}}}}\int {dx{e^{ - \frac{{{{\left( {x - {q_0}} \right)}^2}}}{2} + {\rm{i}}{p_0}x - \frac{{{\rm{i}}{p_0}{q_0}}}{2}}}\left| x \right\rangle }.\label{eq10}
	\end{eqnarray}
where $q_0=\rm{Re}[\sqrt{2}\beta]$ and $p_0=\rm{Im}[\sqrt{2}\beta]$. When the amplitude $|\beta|$ increases, the two basis $\ket{\beta}$ and $\ket{-\beta}$ are quasi-orthogonal. Thus, the state $\ket{\varphi}_{\rm{in}}$ is quite similar to the usual qubit and is called the CV qubit. Our task is to teleport the CV qubit $\ket{\varphi}_{\rm{in}}$ held by Alice onto a DV NV qubit (spin $ 1$) at Bob's site, realizing the distant quantum-state conversion
${\left| \varphi  \right\rangle _{{\rm{in}}}} \mapsto {\left| \psi_T  \right\rangle } = a{\left|  \uparrow  \right\rangle _1} + b{\left|  \downarrow  \right\rangle _1}$.

\begin{figure*}[t]
\includegraphics[scale=0.33]{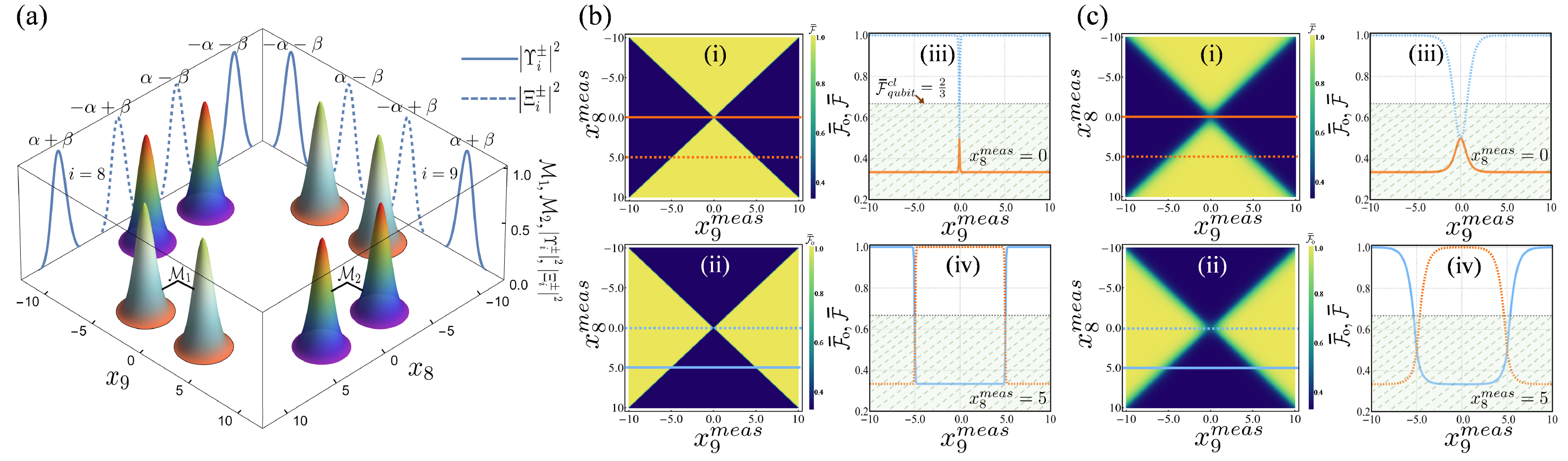}
\caption{(a) Position-space portrait of the measurement results $\mathcal{M}_1$ (green peaks) and $\mathcal{M}_2$ (red peaks). The solid curves (dashed curves) on the wall shows the amplitude of the state $\Upsilon _i^ \pm$ and $\Xi _i^ \pm$. Upper left curves: $|\Upsilon _8^ \pm|^2$ and $|\Xi _8^ \pm|^2$; Upper right curves: $|\Upsilon _9^ \pm|^2$ and $|\Xi _9^ \pm|^2$. (b) Average fidelities $\bar{\mathcal{F}}$ (top row) and $\bar{\mathcal{F}}_o$ (down row) vs HD measurement results $x_8^{\rm{meas}}$ and $x_9^{\rm{meas}}$, respectively, for $\alpha=3$. (iii) and (iv) plot the average fidelities as function of $x_9^{\rm{meas}}$, where the orange-solid and -dotted lines for $\bar{\mathcal{F}}$ and the blue-solid and -dotted lines for $\bar{\mathcal{F}}_o$.
(c) is the same as (b) but with $\alpha=0.2$.}\label{fig3}
\end{figure*}

To implement the teleportation, Bob first prepares his NV center in the superposition state $\ket{\psi_{S1}}=(\ket{\uparrow}_1+\ket{\downarrow}_1)/\sqrt{2}$ and then sends an optical pulse in the coherent state $\ket{\alpha}_0$ to interact with the spin-cavity system $1$. After the reflection by the cavity wall, the state becomes
	\begin{eqnarray}
		\ket{\psi_E}&=&\hat U^{CP}_{10}\left(\ket{\psi_{S1}}\otimes\ket{\alpha}_0\right)\nonumber\\&=&\frac{1}{\sqrt{2}}(\left| \uparrow  \right\rangle_{1} {{\left| \alpha  \right\rangle_4 }}+\left|  \downarrow \right\rangle_{1} {{\left| -\alpha  \right\rangle_4 }}) .\label{eq11}
	\end{eqnarray}
Next, the CV part of the entangled resource Eq. (\ref{eq11}) is sent to Alice via a fiber. To perform the BSM, Alice's mode $4$ is first combined at a symmetric beam splitter with the input mode $5$ to yield the output state
\begin{widetext}
\begin{eqnarray}
{\left| \psi  \right\rangle _{{\rm{out1}}}} &=& {{\hat B}_{45}}\left| {{\psi _E}} \right\rangle  \otimes \left| {{\varphi}} \right\rangle_{{\rm{in}}} \nonumber\\
 &=& \frac{\mathcal{N}}{{\sqrt 2 }}\left( {a{{\left| {\frac{{\alpha  + \beta }}{{\sqrt 2 }}\;} \right\rangle }_6}{{\left| {\frac{{\alpha  - \beta }}{{\sqrt 2 }}\;} \right\rangle }_7} + b{{\left| {\frac{{\alpha  - \beta }}{{\sqrt 2 }}\;} \right\rangle }_6}{{\left| {\frac{{\alpha  + \beta }}{{\sqrt 2 }}\;} \right\rangle }_7}} \right){\left|  \uparrow  \right\rangle _1}\nonumber\\
 &&+ \frac{\mathcal{N}}{{\sqrt 2 }}\left( {a{{\left| {\frac{{ - \alpha  + \beta }}{{\sqrt 2 }}\;} \right\rangle }_6}{{\left| {\frac{{ - \alpha  - \beta }}{{\sqrt 2 }}\;} \right\rangle }_7} + b{{\left| {\frac{{ - \alpha  - \beta }}{{\sqrt 2 }}\;} \right\rangle }_6}{{\left| {\frac{{ - \alpha  + \beta }}{{\sqrt 2 }}\;} \right\rangle }_7}} \right){\left|  \downarrow  \right\rangle _1},\label{eq12}
\end{eqnarray}
where the beam splitter transformation on mode $i$ and $j$ is described by ${{\hat B}_{ij}}=\exp [i({\pi }/{4}\;)(\hat{a}_{i}^{\dagger }{{\hat{a}}_{j}}+\hat{a}_{j}^{\dagger }{{\hat{a}}_{i}})]$ \cite{liao2007near}. In the conventional BSM scheme \cite{cheong2004near}, one needs to set $\alpha=\beta$, then Eq. (\ref{eq12}) can be reexpressed as ${\left| \psi  \right\rangle _{{\rm{out}}}} \propto {\left| {0\;} \right\rangle _6}{\left| {{\rm{even}}} \right\rangle _7}(b{\left|  \uparrow  \right\rangle _1} + a{\left|  \downarrow  \right\rangle _1}) + {\left| {0\;} \right\rangle _6}{\left| {{\rm{odd}}} \right\rangle _7}(b{\left|  \uparrow  \right\rangle _1} - a{\left|  \downarrow  \right\rangle _1}) + {\left| {{\rm{even}}} \right\rangle _6}{\left| {0\;} \right\rangle _7}(a{\left|  \uparrow  \right\rangle _1} + b{\left|  \downarrow  \right\rangle _1}) + {\left| {{\rm{odd}}} \right\rangle _6}{\left| {0\;} \right\rangle _7}(a{\left|  \uparrow  \right\rangle _1} - b{\left|  \downarrow  \right\rangle _1})$, where the even cat state ${\left| {{\rm{even}}} \right\rangle _i} = {\left| {\sqrt 2 \alpha \;} \right\rangle _i} + {\left| { - \sqrt 2 \alpha \;} \right\rangle _i}$ is a superposition of even-number Fock states and the odd cat state ${\left| {{\rm{odd}}} \right\rangle _i} = {\left| {\sqrt 2 \alpha \;} \right\rangle _i} - {\left| { - \sqrt 2 \alpha \;} \right\rangle _i}$ is a superposition of odd-number Fock states. As a result, in the case of large $\alpha$, the four basis $\{{\left| {0\;} \right\rangle _6}{\left| {{\rm{even}}} \right\rangle _7},{\left| {0\;} \right\rangle _6}{\left| {{\rm{odd}}} \right\rangle _7},{\left| {{\rm{even}}} \right\rangle _6}{\left| {0\;} \right\rangle _7},{\left| {{\rm{odd}}} \right\rangle _6}{\left| {0\;} \right\rangle _7}\}$ are nearly orthogonality with each other. Thus, one can discriminate them by detecting the parity of the photon numbers of optical modes $6$ and $7$ \cite{cheong2004near}. For example, if the detected photons in both modes are even and, furthermore, mode 6 detects no photons, then the state of spin 1 would be $b\ket{\uparrow}+a\ket{\downarrow}$, application of a spin-flipping operation completing the quantum teleportation. However, discriminating the parity of photon numbers poses a significant challenge in practice,  particularly for scenarios involving a large number of photons \cite{wang2012photonic}.

Here we propose an alternative approach to realize the BSM, which removes the need for accurate photon-number detection. Instead, it only requires homodyne detection of the light modes and readout of the NV spins.
As shown in Fig. \ref{fig2}, we propose to reflect the light modes $6$ and $7$ by another two auxiliary spin-cavity systems $2$ and $3$, respectively. Similarly, the two spins in cavities are initially prepared in the superposition states $\ket{\psi_{Sk}}=(\ket{\uparrow}_k+\ket{\downarrow}_k)/\sqrt{2}$ with $k\in\{2,3\}$. Then, after the reflection by the cavities the output state reads
\begin{eqnarray}
{\left| \psi  \right\rangle _{{\rm{out2}}}} &=& \hat U_{26}^{CP}\hat U_{37}^{CP}{\left| \psi  \right\rangle _{{\rm{out1}}}} \otimes \left| {{\psi _{S2}}} \right\rangle  \otimes \left| {{\psi _{S3}}} \right\rangle \nonumber\\
 &=& \frac{\mathcal{N}a}{{2\sqrt 2 }}{\left|  \uparrow  \right\rangle _1}\left( {{{\left| {\frac{{\alpha  + \beta }}{{\sqrt 2 }}} \right\rangle }_6}{{\left|  \uparrow  \right\rangle }_2} + {{\left| {\frac{{ - \alpha  - \beta }}{{\sqrt 2 }}} \right\rangle }_6}{{\left|  \downarrow  \right\rangle }_2}} \right)\left( {{{\left| {\frac{{ {\alpha  - \beta } }}{{\sqrt 2 }}} \right\rangle }_7}{{\left|  \uparrow  \right\rangle }_3} + {{\left| {\frac{{ { - \alpha  + \beta } }}{{\sqrt 2 }}} \right\rangle }_7}{{\left|  \downarrow  \right\rangle }_3}} \right)\nonumber\\
 &&+ \frac{\mathcal{N}a}{{2\sqrt 2 }}{\left|  \downarrow  \right\rangle _1}\left( {{{\left| {\frac{{ - \alpha  + \beta }}{{\sqrt 2 }}} \right\rangle }_6}{{\left|  \uparrow  \right\rangle }_2} + {{\left| {\frac{{\alpha  - \beta }}{{\sqrt 2 }}} \right\rangle }_6}{{\left|  \downarrow  \right\rangle }_2}} \right)\left( {{{\left| {\frac{{ { - \alpha  - \beta } }}{{\sqrt 2 }}} \right\rangle }_7}{{\left|  \uparrow  \right\rangle }_3} + {{\left| {\frac{{{\alpha  + \beta } }}{{\sqrt 2 }}} \right\rangle }_7}{{\left|  \downarrow  \right\rangle }_3}} \right)\nonumber\\
 &&+ \frac{\mathcal{N}b}{{2\sqrt 2 }}{\left|  \uparrow  \right\rangle _1}\left( {{{\left| {\frac{{\alpha  - \beta }}{{\sqrt 2 }}} \right\rangle }_6}{{\left|  \uparrow  \right\rangle }_2} + {{\left| {\frac{{ - \alpha  + \beta }}{{\sqrt 2 }}} \right\rangle }_6}{{\left|  \downarrow  \right\rangle }_2}} \right)\left( {{{\left| {\frac{{\alpha  + \beta }}{{\sqrt 2 }}} \right\rangle }_7}{{\left|  \uparrow  \right\rangle }_3} + {{\left| {\frac{{ - \alpha  - \beta }}{{\sqrt 2 }}} \right\rangle }_7}{{\left|  \downarrow  \right\rangle }_3}} \right)\nonumber\\
 &&+ \frac{\mathcal{N}b}{{2\sqrt 2 }}{\left|  \downarrow  \right\rangle _1}\left( {{{\left| {\frac{{ - \alpha  - \beta }}{{\sqrt 2 }}} \right\rangle }_6}{{\left|  \uparrow  \right\rangle }_2} + {{\left| {\frac{{\alpha  + \beta }}{{\sqrt 2 }}} \right\rangle }_6}{{\left|  \downarrow  \right\rangle }_2}} \right)\left( {{{\left| {\frac{{ - \alpha  + \beta }}{{\sqrt 2 }}} \right\rangle }_7}{{\left|  \uparrow  \right\rangle }_3} + {{\left| {\frac{{\alpha  - \beta }}{{\sqrt 2 }}} \right\rangle }_7}{{\left|  \downarrow  \right\rangle }_3}} \right).\label{eq13}
\end{eqnarray}
Furthermore, before the detections two microwave (MW) ${\pi}/{2}$-pulse that drive the transition $\ket{\uparrow}_k\leftrightarrow\ket{\downarrow}_k$ are applied to the spin states to realize the transformations $\ket{\uparrow}_k\mapsto\ket{\uparrow}_k+\ket{\downarrow}_k$, $\ket{\downarrow}_k\mapsto\ket{\uparrow}_k-\ket{\downarrow}_k$, which yields the final state
\begin{eqnarray}
{\left| \psi  \right\rangle _{\rm{out3}}} &=& \frac{\mathcal{N}}{{2\sqrt 2 }}\left[ {\left( {{{\left|  \uparrow  \right\rangle }_2}{{\left|  \uparrow  \right\rangle }_3}\left| {\Upsilon _8^ + } \right\rangle \left| {\Xi _9^ + } \right\rangle  + {{\left|  \downarrow  \right\rangle }_2}{{\left|  \downarrow  \right\rangle }_3}\left| {\Upsilon _8^ - } \right\rangle \left| {\Xi _9^ - } \right\rangle } \right)\left( {a{{\left|  \uparrow  \right\rangle }_1} + b{{\left|  \downarrow  \right\rangle }_1}} \right)} \right.\nonumber\\
 &&+ \left( {{{\left|  \uparrow  \right\rangle }_2}{{\left|  \downarrow  \right\rangle }_3}\left| {\Upsilon _8^ + } \right\rangle \left| {\Xi _9^ - } \right\rangle  + {{\left|  \downarrow  \right\rangle }_2}{{\left|  \uparrow  \right\rangle }_3}\left| {\Upsilon _8^ - } \right\rangle \left| {\Xi _9^ + } \right\rangle } \right)\left( {a{{\left|  \uparrow  \right\rangle }_1} - b{{\left|  \downarrow  \right\rangle }_1}} \right)\nonumber\\
 &&+ \left( {{{\left|  \uparrow  \right\rangle }_2}{{\left|  \uparrow  \right\rangle }_3}\left| {\Xi _8^ + } \right\rangle \left| {\Upsilon _9^ + } \right\rangle  + {{\left|  \downarrow  \right\rangle }_2}{{\left|  \downarrow  \right\rangle }_3}\left| {\Xi _8^ - } \right\rangle \left| {\Upsilon _9^ - } \right\rangle } \right)\left( {a{{\left|  \downarrow  \right\rangle }_1} + b{{\left|  \uparrow  \right\rangle }_1}} \right)\nonumber\\
&&\left. { - \left( {{{\left|  \uparrow  \right\rangle }_2}{{\left|  \downarrow  \right\rangle }_3}\left| {\Xi _8^ + } \right\rangle \left| {\Upsilon _9^ - } \right\rangle  + {{\left|  \downarrow  \right\rangle }_2}{{\left|  \uparrow  \right\rangle }_3}\left| {\Xi _8^ - } \right\rangle \left| {\Upsilon _9^ + } \right\rangle } \right)\left( {a{{\left|  \downarrow  \right\rangle }_1} - b{{\left|  \uparrow  \right\rangle }_1}} \right)} \right],\label{eq14}
\end{eqnarray}
where we have defined the cat states $\Upsilon _i^ \pm  = {\left| {(\alpha  + \beta )/\sqrt 2 } \right\rangle _i} \pm {\left| { - (\alpha  + \beta )/\sqrt 2 } \right\rangle _i}$ and $\Xi _i^ \pm  = {\left| {(\alpha  - \beta )/\sqrt 2 } \right\rangle _i} \pm {\left| { - (\alpha  - \beta )/\sqrt 2 } \right\rangle _i}$.
\end{widetext}
Eq. (\ref{eq14}) is the main result of this work, showing that the target qubits (that is, the last term in bracket on each line) get entangled with the auxiliary spin qubits $2,3$ and light modes $8,9$. To distill the quantum information from Eq. (\ref{eq14}), measurements on these auxiliary quantum systems are required. For spins, projection measurements on (or readout of) the NV spins are necessary, which, in practise, are usually performed via resonance fluorescence \cite{jayakumar2018spin,ju2021preparations,irber2021robust,shields2015efficient}. Recently, it has been shown that high-fidelity ($>95\%$) single-shot readout of single NV center in diamond is possible \cite{zhang2021high} via spin-to-charge conversion \cite{giri2023charge}. For light, we assume $\alpha,\beta$ to be real, then the amplitudes of the states $\Upsilon _i^ \pm$ and $\Xi _i^ \pm$ in the $x$ basis [according to Eq. (\ref{eq10})] are
\begin{eqnarray}
{\left| {\Upsilon _i^ \pm } \right|^2} \propto {e^{ - \left[ {x -  {\mathop{}\nolimits} \left( {\alpha  + \beta } \right)} \right]^2}} + {e^{ - \left[ {x + {\mathop{}\nolimits} \left( {\alpha  + \beta } \right)} \right]^2}},\label{eq15}\\
{\left| {\Xi _i^ \pm } \right|^2} \propto {e^{ - \left[ {x - {\mathop{}\nolimits} \left( {\alpha  - \beta } \right)} \right]^2}} + {e^{ - \left[ {x + {\mathop{}\nolimits} \left( {\alpha  - \beta } \right)} \right]^2}}.\label{eq16}
\end{eqnarray}
Obviously, in the case of $\mid\alpha+\beta\mid\gg\mid\alpha-\beta\mid\gg 0$, the distribution of ${\left| {\Upsilon _i^ \pm } \right|^2},{\left| {\Xi _i^ \pm } \right|^2}$ should be four distinct peaks, as shown in \cref{fig3}(a). In other words, the four cat states $\{\ket{{\Xi _i^ \pm }},\ket{{\Upsilon _i^ \pm }}\}$ are nearly orthogonal to each other. As a result, one may discriminate the quantum states $\ket{{\Upsilon _i^ \pm }}$ from $\ket{{\Xi _i^ \pm }}$ with high probability via only HD detection of the position quadratures of the light modes $8$ and $9$. More specifically, if the detected result $x^{\rm{meas}}$ is around the centers $\pm(\alpha-\beta)$, then the states should be $\ket{{\Xi _i^ \pm }}$, and, on the contrary, if $x^{\rm{meas}}$ is around the centers $\pm(\alpha+\beta)$, then the states would be $\ket{{\Upsilon _i^ \pm }}$.

As an example, if the spins measurement outcomes are $\ket{\uparrow}_2\ket{\uparrow}_3$, then, according to Eq. (\ref{eq14}), the projected state would be
\begin{eqnarray}
{\left| \psi  \right\rangle _{{p}}} &\propto& \left| {\Upsilon _8^ + } \right\rangle \left| {\Xi _9^ + } \right\rangle \left( {a{{\left|  \uparrow  \right\rangle }_1} + b{{\left|  \downarrow  \right\rangle }_1}} \right)\nonumber\\
 &&+ \left| {\Xi _8^ + } \right\rangle \left| {\Upsilon _9^ + } \right\rangle \left( {a{{\left|  \downarrow  \right\rangle }_1} + b{{\left|  \uparrow  \right\rangle }_1}} \right).\label{eq17}
\end{eqnarray}
Next, a HD detection of the position quadrature of mode $8$ is performed, which collapses the light state into a position state $\ket{x^{\rm{meas}}_8}$. As analyzed previously, there are high probabilities of finding the measurement results around the four coherent amplitudes $\pm(\alpha\pm\beta)$ [see \cref{fig3}(a)]. Depending on which coherent amplitudes the mode $8$ collapses into, the measurement of the mode $9$ yields two types of results: if $x^{\rm{meas}}_8$ is approximately at $\alpha+\beta$, then measurement outcome $x^{\rm{meas}}_9$ would be approximately at either the amplitude $\alpha -\beta$ or $-(\alpha-\beta)$, according to Eq. (\ref{eq17}). As a result, the outcome of the HD detections would be eight distinct Gaussian peaks, as shown in \cref{fig3}(a), and the measurement processes finally project the NV spin $1$ onto
\begin{eqnarray}
{\left| \psi  \right\rangle _f} = {\mathcal{N}_f}\left[ {\left( {a{{\mathcal{M}}_1} + b{{\mathcal{M}}_2}} \right){{\left|  \uparrow  \right\rangle }_1} + \left( {b{{\mathcal{M}}_1} + a{{\mathcal{M}}_2}} \right){{\left|  \downarrow  \right\rangle }_1}} \right],\label{eq18}\nonumber\\
\end{eqnarray}
where the measurement results $\mathcal{M}_{1,2}$ and the normalization constant $\mathcal{N}_f$ are given by
\begin{eqnarray}
{\mathcal{M}_1} &=& \left\langle {{x_8^{{\rm{meas}}}}}
 \mathrel{\left | {\vphantom {{x_8^{{\rm{meas}}}} {\Upsilon _8^ + }}}
 \right. \kern-\nulldelimiterspace}
 {{\Upsilon _8^ + }} \right\rangle \left\langle {{x_9^{{\rm{meas}}}}}
 \mathrel{\left | {\vphantom {{x_9^{{\rm{meas}}}} {\Xi _9^ + }}}
 \right. \kern-\nulldelimiterspace}
 {{\Xi _9^ + }} \right\rangle , \nonumber\\
{\mathcal{M}_2} &=& \left\langle {{x_8^{{\rm{meas}}}}}
 \mathrel{\left | {\vphantom {{x_8^{{\rm{meas}}}} {\Xi _8^ + }}}
 \right. \kern-\nulldelimiterspace}
 {{\Xi _8^ + }} \right\rangle \left\langle {{x_9^{{\rm{meas}}}}}
 \mathrel{\left | {\vphantom {{x_9^{{\rm{meas}}}} {\Upsilon _9^ + }}}
 \right. \kern-\nulldelimiterspace}
 {{\Upsilon _9^ + }} \right\rangle ,\nonumber\\
{\mathcal{N}_f} &=& \left[\mathcal{M}_1^2 + \mathcal{M}_2^2 + 2\left( {a{b^*} + b{a^*}} \right){\mathcal{M}_1}{\mathcal{M}_2}\right]^{-1/2}.\nonumber
\end{eqnarray}
From Eq. (\ref{eq18}) we see immediately that if the measurement results satisfy $\mathcal{M}_1\gg\mathcal{M}_2$ then the spin state would be exactly the target state $\ket{\psi_T}$, showing that Alice's CV-encoded quantum state has been successfully teleported onto Bob's DV spin. On the contrary, if the measurement results fulfill $\mathcal{M}_1\ll\mathcal{M}_2$ then the resulting state is ${\left| \psi  \right\rangle _{f}} \simeq b{\left|  \uparrow  \right\rangle _1} + a{\left|  \downarrow  \right\rangle _1}$, and, after applying a bit flip operation described by ${\hat U_o} = \left|  \uparrow  \right\rangle _{\kern-0.1em 11 \kern-0.2em}\left\langle  \downarrow  \right| + \left|  \downarrow  \right\rangle _{\kern-0.1em 11 \kern-0.2em}\left\langle  \uparrow  \right|$, Bob gets the state $\hat U_o\ket{\psi}_{f}=a{\left|  \uparrow  \right\rangle _1} + b{\left|  \downarrow  \right\rangle _1}$, which is exactly the same as $\ket{\psi_T}$.
Despite being very small, there is still a probability that the two measurement results, $\mathcal{M}_1$ and $\mathcal{M}_2$, are comparable. In these cases, the teleported states are imperfect.
 To quantify how well the quantum state was teleproted, we utilize the fidelity $\mathcal{F} = {\left\langle {\psi_T \left| {{\hat\rho _{f}}} \right|\psi_T } \right\rangle }$, which measures the overlap between the target state $\ket{\psi_T}$ and the
density operator  $\hat\rho_f$ for a teleported state \cite{furusawa1998unconditional}. Using Eq. (\ref{eq18}) we find the teleportation fidelity of the form
\begin{eqnarray}
\mathcal{F} = \mathcal{N}_f^2{\left[ {{{\mathcal{M}}_1} + \left(  {a{b^*} + b{a^*}} \right){{\mathcal{M}}_2}} \right]^2}.\label{eq19}
\end{eqnarray}
Next, the average fidelity will be calculated by setting $a=\cos(\theta/2)$ and $b=\sin(\theta/2)e^{i\phi}$ \cite{muschik2006efficient},
and after integrating over the whole Bloch sphere, we obtain
\begin{eqnarray}
\bar {\mathcal{F}} &=& \frac{1}{{4\pi }}\int_0^\pi  {d\theta } \int_0^{2\pi } {d\phi \mathcal{F}} \nonumber\\
 &=& 1 - \frac{1}{{4\pi }}\int_0^\pi  {d\theta } \int_0^{2\pi } {d\phi \frac{{\sin \theta  - {{\sin }^3}\theta {{\cos }^2}\phi }}{{1 + 2\varsigma \sin \theta \cos \phi  + {\varsigma ^2}}}} \nonumber\\
 &\simeq& \left\{ {\begin{array}{*{20}{c}}
{1}~~&\left({\varsigma} \to \infty\right),\\
{\frac{1}{3}}~~&\left({\varsigma} \to 0\right),~
\end{array}} \right.\label{eq20}
\end{eqnarray}
where $\varsigma=\mathcal{M}_1/\mathcal{M}_2$.
\begin{table}[t]
\renewcommand{\arraystretch}{1.2}
\caption{All possible measurement outcomes obtained at Alice's site, and the corresponding spin state at Bob's site, which will be transformed to the target state $\ket{\psi_T}$ via the unitary operation $\hat U_o$.}
\label{table1}
\begin{ruledtabular}
\begin{tabular}{llc}
\textrm{Alice's measurement results} &
\textrm{Bob's states} &
\textrm{$\hat U_o$} \\
\hline
$\begin{aligned}{~~\left|  \uparrow  \right\rangle _2}{\left|  \uparrow  \right\rangle _3}\left| {\Upsilon _8^ + } \right\rangle \left| {\Xi _9^ + } \right\rangle\\ {\left|  \downarrow  \right\rangle _2}{\left|  \downarrow  \right\rangle _3}\left| {\Upsilon _8^ - } \right\rangle \left| {\Xi _9^ - } \right\rangle\end{aligned}$ & ${a{{\left|  \uparrow  \right\rangle }_1} + b{{\left|  \downarrow  \right\rangle }_1}}$ & $\bigl( \begin{smallmatrix}1 &0\\ 0 &1\end{smallmatrix}\bigr)$ \\
\hline
$\begin{aligned}{{{~~\left|  \uparrow  \right\rangle }_2}{{\left|  \downarrow  \right\rangle }_3}\left| {\Upsilon _8^ + } \right\rangle \left| {\Xi _9^ - } \right\rangle }\\{{{\left|  \downarrow  \right\rangle }_2}{{\left|  \uparrow  \right\rangle }_3}\left| {\Upsilon _8^ - } \right\rangle \left| {\Xi _9^ + } \right\rangle }\end{aligned}$ &${a{{\left|  \uparrow  \right\rangle }_1} - b{{\left|  \downarrow  \right\rangle }_1}}$& $\bigl( \begin{smallmatrix}1 &0\\ 0 &-1\end{smallmatrix}\bigr)$ \\
\hline
$\begin{aligned}{{{~~\left|  \uparrow  \right\rangle }_2}{{\left|  \uparrow  \right\rangle }_3}\left| {\Xi _8^ + } \right\rangle \left| {\Upsilon _9^ + } \right\rangle }\\ {{{\left|  \downarrow  \right\rangle }_2}{{\left|  \downarrow  \right\rangle }_3}\left| {\Xi _8^ - } \right\rangle \left| {\Upsilon _9^ - } \right\rangle }\end{aligned}$ & ${a{{\left|  \downarrow  \right\rangle }_1} + b{{\left|  \uparrow  \right\rangle }_1}}$ & $\bigl( \begin{smallmatrix}0 &1\\ 1 &0\end{smallmatrix}\bigr)$ \\
\hline
$\begin{aligned}{{{~~\left|  \uparrow  \right\rangle }_2}{{\left|  \downarrow  \right\rangle }_3}\left| {\Xi _8^ + } \right\rangle \left| {\Upsilon _9^ - } \right\rangle }\\{{{\left|  \downarrow  \right\rangle }_2}{{\left|  \uparrow  \right\rangle }_3}\left| {\Xi _8^ - } \right\rangle \left| {\Upsilon _9^ + } \right\rangle }\end{aligned}$ & ${a{{\left|  \downarrow  \right\rangle }_1} - b{{\left|  \uparrow  \right\rangle }_1}}$ & $\bigl( \begin{smallmatrix}0 &1\\ -1 &0\end{smallmatrix}\bigr)$ \\
\end{tabular}
\end{ruledtabular}
\end{table}
As expected, near-unity teleportation fidelity can be achieved when $\mathcal{M}_1$ is much larger than $\mathcal{M}_2$. For the case of small $\varsigma$, the fidelity should turn out to be $\mathcal{F}_o =\bra{\psi_T}\hat U_o\hat\rho _{f}\hat U_o^\dag\ket{\psi_T} $. Fig.  \ref{fig3}(b)(i) and (ii) show the average fidelities as a function of measurement results of $x_8^{\rm{meas}}$ and $x_9^{\rm{meas}}$. One can see clearly that the fidelities $\bar{\mathcal{F}}$ and $\bar{\mathcal{F}}_o$ complement each other. As a result, near-deterministic teleportation between a CV optical qumode and a DV spin qubit can be achieved. There are, however, small probabilities that the scheme will fail, that is, as shown in Fig. \ref{fig3}(b)(iii) and (iv), near the regime of diagonal lines (where $|x_8^{\rm{meas}}|=|x_9^{\rm{meas}}|$) the fidelities are lower than the maximal average fidelity $\bar{\mathcal{F}}^{cl}_{\rm{qubit}}$ that can be achieved for qubit state by a classical strategy \cite{popescu1994bell,massar1995optimal,pirandola2015advances,derka1998universal}. As the coherent amplitudes, $\alpha$ and $\beta$, decrease, the unfavorable low-fidelity regimes will be expanded, as shown in Fig. \ref{fig3}(c). Conversely, in the case of large coherent amplitudes, the low-fidelity regimes will be compressed into two lines along the diagonal. The probabilities to find the measurement results lying on the two diagonal lines are
\begin{eqnarray}
{\mathcal{P}_{{\rm{fail}}}} &\propto&\left\langle {x}
\mathrel{\left | {\vphantom {x {\Upsilon _8^ + }}}
	\right. \kern-\nulldelimiterspace}
{{\Upsilon _8^ + }} \right\rangle \left\langle {x}
\mathrel{\left | {\vphantom {x {\Xi _9^2}}}
	\right. \kern-\nulldelimiterspace}
{{\Xi _9^+}} \right\rangle\nonumber\\
&\propto& {e^{ - {\beta ^2}}}\left[ {{e^{ - {{\left( {x - 3\beta } \right)}^2}}} + {e^{ - {{\left( {x + 3\beta } \right)}^2}}}} \right]\nonumber\\
&&+ {e^{ - 9{\beta ^2}}}\left[ {{e^{ - {{\left( {x - \beta } \right)}^2}}} + {e^{ - {{\left( {x + \beta } \right)}^2}}}} \right],\label{eq21}
\end{eqnarray}
where we have assumed $x_8^{\rm{meas}}=x_9^{\rm{meas}}\equiv x$ and, for simplicity, $\alpha=3\beta$. From Eq. (\ref{eq21}) we can see that the failure probabilities decay exponentially with the coherent amplitude $\beta$, as expected. Consequently, our proposed scheme can work deterministically under the condition of large coherent amplitude ($\beta\gg 1$). Table \ref{table1} also shows other possible measurement outcomes.
As illustrated in Fig. \ref{fig2}, Alice will transmit her measurement outcomes to Bob through a classical channel. Based on the received information, Bob applies a unitary transformation $\hat U_o$ to spin $1$ to complete the teleportation process.

\section{\label{sec:level2.3} The noise effect}	
Up to now, we have neglected the noise effects. Considering the fact that one can perform high-fidelity readout of the NV center \cite{zhang2021high} and high-efficiency HD measurement of light \cite{wenger2003maximal} in reality, we neglect the noise effects imposed by the detections.
Here, we mainly concentrate on the decoherences that occur during entanglement distribution \cite{yin2017satellite}. For spin, it is inevitable that it undergoes spin dephasing and spin relaxation during the time interval $\tau$ when the optical pulse propagates in the fiber, which can be described using a master equation \cite{groszkowski2022reservoir}: $\dot {\hat \rho}  ={{{\gamma _\phi }}}{\cal D}[ {{{\hat \sigma }_z}} ]\hat \rho  + {\gamma}{\cal D}[ {{{\hat \sigma }_ - }} ]\hat \rho $, where ${\cal D}\left[ {\hat c} \right]\hat\rho  = \hat c\hat\rho {{\hat c}^\dag } - \frac{1}{2}\left( {{{\hat c}^\dag }\hat c\hat\rho  + \hat\rho {{\hat c}^\dag }\hat c} \right)$, and $\gamma$ ($=T_1^{-1}$), $\gamma_\phi$ ($=T_2^{-1}$) denote the dephasing and relaxation rates, respectively. For light, propagation in the fiber will cause photon loss, which can be modeled by a beam splitter type admixture of the optical mode with vacuum components \cite{fiuravsek2006single,hammerer2005teleportation}. After taking into account all these realistic noises, the distributed spin-light entanglement becomes
\begin{eqnarray}
	{\hat\rho(\tau)} &=& \frac{1}{2}\left[ \left|  \uparrow  \right\rangle _{\kern-0.1em 11 \kern-0.2em}\left\langle  \uparrow  \right| \otimes \left| {t\alpha } \right\rangle _{\kern-0.1em 44 \kern-0.2em}\left\langle {t\alpha } \right| \otimes \left| {r\alpha } \right\rangle _{\kern-0.1em vv \kern-0.2em}\left\langle {r\alpha } \right| \right.\nonumber\\
	&&+ {e^{ - \frac{{{(4\gamma _\phi+\gamma) }\tau }}{2}}}\left|  \uparrow  \right\rangle _{\kern-0.1em 11 \kern-0.2em}\left\langle  \downarrow  \right| \otimes \left| {t\alpha } \right\rangle _{\kern-0.1em 44 \kern-0.2em}\left\langle { - t\alpha } \right| \otimes \left| {r\alpha } \right\rangle _{\kern-0.1em vv \kern-0.2em}\left\langle { - r\alpha } \right|\nonumber\\
	&&+ {e^{ - \frac{{{(4\gamma_\phi+\gamma)}\tau }}{2}}}\left|  \downarrow  \right\rangle _{\kern-0.1em 11 \kern-0.2em}\left\langle  \uparrow  \right| \otimes \left| { - t\alpha } \right\rangle _{\kern-0.1em 44 \kern-0.2em}\left\langle {t\alpha } \right| \otimes \left| { - r\alpha } \right\rangle _{\kern-0.1em vv \kern-0.2em}\left\langle {r\alpha } \right|\nonumber\\
	&&+ {e^{ - {\gamma}\tau }}\left|  \downarrow  \right\rangle _{\kern-0.1em 11 \kern-0.2em}\left\langle  \downarrow  \right| \otimes \left| { - t\alpha } \right\rangle _{\kern-0.1em 44 \kern-0.2em}\left\langle { - t\alpha } \right| \otimes \left| { - r\alpha } \right\rangle _{\kern-0.1em vv \kern-0.2em}\left\langle { - r\alpha } \right|\nonumber\\
	&&+ \left( {1 - {e^{ - {\gamma}\tau }}} \right)\left|  \uparrow  \right\rangle _{\kern-0.1em 11 \kern-0.2em}\left\langle  \uparrow  \right| \otimes \left| { - t\alpha } \right\rangle _{\kern-0.1em 44 \kern-0.2em}\left\langle { - t\alpha } \right| \nonumber\\&&\left.\otimes \left| { - r\alpha } \right\rangle _{\kern-0.1em vv \kern-0.2em}\left\langle { - r\alpha } \right|\right],\label{eq22}
\end{eqnarray}
where $\ket{\cdot}_v$ denotes the auxiliary vacuum mode, and the last term arises because of quantum jump. The overall photon loss rate $r=\sqrt{1-\eta_p}$ ($t=\sqrt{\eta_p}$) scales exponentially with the communication distance $d_0=c\tau$ with $c$ the velocity of light, where the overall efficiency $\eta_p= e^{-d_0/L_{\rm{att}}}$ with $L_{\rm{att}}$ being the channel attenuation
length \cite{duan2001long,van2011optical}. Using Eq. (\ref{eq22}) the teleportation protocol is done along the same lines outlined in the section above. Finally, we obtain the teleported state in the presence of noises (for more details see Appendix A)
      \begin{figure}[t]
\includegraphics[scale=0.35]{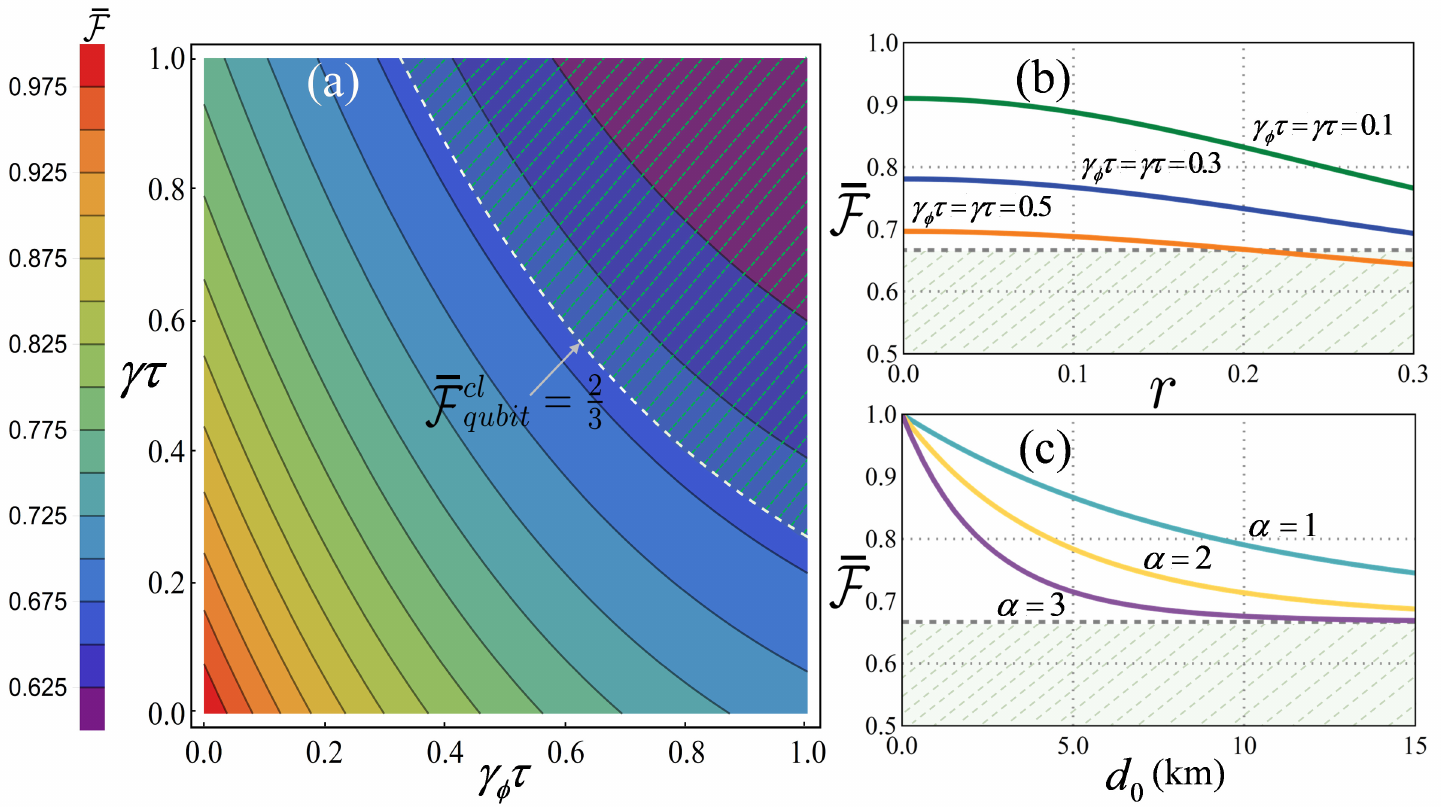}
\caption{(a) Average fidelity for hybrid quantum teleportation vs spin dephasing $\gamma_\phi\tau$ and spin relaxation $\gamma\tau$ for $r\alpha=0.03$. The atomic decays in the shaded area yields a fidelity that can be achieved by a classical strategy. (b) Average fidelity achievable vs light loss $r$ in the presence of spin decays $\gamma_\phi\tau=\gamma\tau=0.1,0.3,0.5$. (c) Average fidelity vs teleportation distance $d_0$ for coherent amplitude (of the quantum-channel pulse) $\alpha=1,2,3$ , spin dephasing $\gamma_\phi=10$ kHz, spin relaxation $\gamma=0$, and light attenuation length $L_{\rm{att}}=25.5$ km. The fidelity benchmark is $2/3$ (dashed line) in (b) and (c).
 }\label{fig4}
\end{figure}

	\begin{eqnarray}
\hat{\rho}_f&=&\left[ \left| a \right|^2+\left| b \right|^2\left( 1-e^{-\gamma \tau} \right) \right] \left| \uparrow \right. \rangle _{\kern -0.1em 11\kern -0.2em}\left. \langle \uparrow \right|
\nonumber\\
&&+ab^*e^{-\frac{\left( 4\gamma _{\phi}+\gamma \right) \tau}{2}-r^2\alpha ^2}\left| \uparrow \right. {\rangle _{\kern -0.1em 11\kern -0.2em}}\left. \langle \downarrow \right|
\nonumber\\
&&+ba^*e^{-\frac{\left( 4\gamma _{\phi}+\gamma \right) \tau}{2}-r^2\alpha ^2}\left| \downarrow \right. \rangle _{\kern -0.1em 11\kern -0.2em}\left. \langle \uparrow \right|
\nonumber\\
&&+\left| b \right|^2e^{-\gamma \tau}\left| \downarrow \right. \rangle _{\kern -0.1em 11\kern -0.2em}\left. \langle \downarrow \right|.\label{eq23}
	\end{eqnarray}
In deriving Eq. (\ref{eq23}) we have assumed that the measurement results are $
\left| \uparrow \right. \rangle _2\left| \uparrow \right. \rangle _3\left| \Upsilon _{8}^{+} \right. \rangle \left| \Xi _{9}^{+} \right. \rangle$. The final state $\hat{\rho}_f$ suggests that all the noise effects contribute to the decoherence of the NV spin, while only spin relaxation causes the population transfer from the state $\left|  \downarrow  \right\rangle $ to the state $\left|\uparrow \right\rangle $. From Eq. (\ref{eq23}) one may directly calculate the average fidelity and obtain
	\begin{eqnarray}
\bar{\mathcal{F}} = \frac{1}{2} + \frac{1}{6}{e^{ - \gamma \tau }} + \frac{1}{3}{e^{ - \frac{{\left( {4{\gamma _\phi } + \gamma } \right)\tau }}{2} - {r^2}{\alpha ^2}}}.\label{eq24}
	\end{eqnarray}
Figure \ref{fig4}(a) plots the fidelities as a function of $\gamma_\phi\tau$ and $\gamma\tau$, showing that the protocol is more susceptible to spin dephasing. As shown in Fig. \ref{fig4}(b), for values of $\gamma_\phi\tau,\gamma\tau,r\leq 0.3$, the average fidelity is still well above the classical bound on the fidelity.

In addition to teleportation fidelity, teleportation distance is also a crucial factor for the implementation of quantum networks and should be evaluated. For NV spin, the single spin relaxation $T_1^{-1}$ is usually much smaller than the single spin dephasing $T_2^{-1}$ \cite{doherty2013nitrogen}, especially at low temperatures as $T_1$ can be several minutes \cite{bennett2013phonon}. Thus, one may safely ignore the spin relaxation ($\gamma\simeq 0$). Taking experimental feasible values of  $\gamma_\phi\simeq 10$ kHz at room temperature \cite{doherty2013nitrogen} and $L_{\rm{att}}\simeq 25.5$ km (corresponding
to an optical fiber absorption of $0.17$ dB/km \cite{van2011optical}), we are able to plot in Fig. \ref{fig4}(c) the average fidelity as a function of teleportation distance $d_0$, showing that the fidelity larger than $0.9$ is still obtainable for $d_0\approx3$ km and, moreover, the fidelity remains significantly above the classical bound when $d_0=15$ km and $\alpha\leq 2$. For low temperature NV spin, the spin dephasing can be of the order of $100$ Hz \cite{bennett2013phonon}, which is much smaller than the light attenuation rate $c/L_{\rm{att}}\approx 10$ kHz. Then, the dominant noise will be the light loss, leading to the average fidelity $\bar{\mathcal{F}} = \frac{2}{3} +  \frac{1}{3}{e^{ - {r^2}{\alpha ^2}}}$. To achieve a high fidelity of $\bar{\mathcal{F}}_h$, it is necessary that ${d_0} \ge \frac{{{L_{\rm{att}}}}}{\alpha^2 } \ln {\left( {3{\bar{\mathcal{F}}_h} - 2} \right)} $, indicating that long attenuation length or small coherent amplitude is required for achieving high-fidelity and long-distance quantum teleportation.

\section{\label{sec:level3} Conclusion}
In conclusion, we have presented a simple and realistic protocol for deterministic teleportation of a flying qumode, carried by an optical pulse, onto a stationary qubit, encoded in an NV center contained in an optical cavity. The scheme relies on a hybrid entanglement between the qumode and the qubit, which can be created by simply reflecting a coherent light pulse from the spin-cavity system, inducing a CP rotation of the field mode depending on the state of the NV center. By adding two ancillary NV spins and utilizing their CP interactions with the sender's field modes, we can achieve the performance of BSM solely through the HD detection of the light modes and the measurement of the NV-spin directions, which greatly simplifies the experimental implementation since neither photon-number parity measurement \cite{lee2013near} nor nonlinear Kerr interaction \cite{liao2006new} is involved.
An estimation of the noise effects indicates that the proposed scheme enables surpassing the maximum average fidelity achievable for qubit states by a classical strategy under realistic experimental conditions, and long attenuation length or small coherent amplitude is necessary to achieve long-distance teleportation. Our proposed scheme can also be extended to other spin systems, such as a neutral atom in cavity \cite{duan2005robust}.

	\begin{acknowledgments}
		This work was supported by the Innovation Program for Quantum Science and Technology (2023ZD0300904), the Natural Science Foundation of China (Grants No. 22273067), the Natural Science Foundation of Zhejiang province, China (Grant No. LQ23A040001, No. LY24A040004), and the Department of Education of Zhejiang Province, China (Grant No. Y202146469).		
	\end{acknowledgments}
   \vspace{0.11cm}
\appendix
	\section{Consideration of noise in the teleportation protocol}
\label{App:a}
	In this Appendix, we give the details of the derivation of Eq.~(\ref{eq23}). The spin-light entangled state of Eq.~(\ref{eq22}) is admixed with the optical state $\ket{\psi}_{\rm{in}}$ to be teleportated at $B{S_1}$, yielding
\begin{widetext}
	\begin{eqnarray}
\hat\rho_{\rm{out1}}&=&{\hat B}_{45}\hat\rho(\tau)\otimes\ket{\psi}_{\kern-0.1em \rm{in}~\rm{in} \kern-0.2em}\bra{\psi}{\hat B}_{45}^\dag\nonumber\\
	&=&\frac{1}{2}{{\left| a \right|}^{2}}\left| \uparrow  \right\rangle _{\kern-0.1em 11 \kern-0.2em}\left\langle  \uparrow  \right|\otimes \left| r\alpha  \right\rangle _{\kern-0.1em vv \kern-0.2em}\left\langle  r\alpha  \right|\otimes \left| {{\Gamma }_{+}} \right\rangle _{\kern-0.1em 66 \kern-0.2em}\left\langle  {{\Gamma }_{+}} \right|\otimes \left| {{\Gamma }_{-}} \right\rangle _{\kern-0.1em 77 \kern-0.2em}\left\langle  {{\Gamma }_{-}} \right| \nonumber\\
		&& +{{\left| a \right|}^{2}}{{e}^{-\frac{\left( \gamma +4{{\gamma }_{\phi }} \right)\tau }{2}}}\left| \uparrow  \right\rangle _{\kern-0.1em 11 \kern-0.2em}\left\langle  \downarrow  \right|\otimes \left| r\alpha  \right\rangle _{\kern-0.1em vv \kern-0.2em}\left\langle  -r\alpha  \right|\otimes \left| {{\Gamma }_{+}} \right\rangle _{\kern-0.1em 66 \kern-0.2em}\left\langle  -{{\Gamma }_{-}} \right|\otimes \left| {{\Gamma }_{-}} \right\rangle _{\kern-0.1em 77 \kern-0.2em}\left\langle  -{{\Gamma }_{+}} \right| \nonumber\\
		&& +\frac{1}{2}{{\left| a \right|}^{2}}{{e}^{-\gamma \tau }}\left| \downarrow  \right\rangle _{\kern-0.1em 11 \kern-0.2em}\left\langle  \downarrow  \right|\otimes \left| -r\alpha  \right\rangle _{\kern-0.1em vv \kern-0.2em}\left\langle  -r\alpha  \right|\otimes \left| -{{\Gamma }_{-}} \right\rangle _{\kern-0.1em 66 \kern-0.2em}\left\langle  -{{\Gamma }_{-}} \right|\otimes \left| -{{\Gamma }_{+}} \right\rangle _{\kern-0.1em 77 \kern-0.2em}\left\langle  -{{\Gamma }_{+}} \right| \nonumber\\
		&& +\frac{1}{2}{{\left| a \right|}^{2}}\left( 1-{{e}^{-\gamma \tau }} \right)\left| \uparrow  \right\rangle _{\kern-0.1em 11 \kern-0.2em}\left\langle  \uparrow  \right|\otimes \left| -r\alpha  \right\rangle _{\kern-0.1em vv \kern-0.2em}\left\langle  -r\alpha  \right|\otimes \left| -{{\Gamma }_{-}} \right\rangle _{\kern-0.1em 66 \kern-0.2em}\left\langle  -{{\Gamma }_{-}} \right|\otimes \left| -{{\Gamma }_{+}} \right\rangle _{\kern-0.1em 77 \kern-0.2em}\left\langle  -{{\Gamma }_{+}} \right| \nonumber\\
		&& +a{{b}^{*}}\left| \uparrow  \right\rangle _{\kern-0.1em 11 \kern-0.2em}\left\langle  \uparrow  \right|\otimes \left| r\alpha  \right\rangle _{\kern-0.1em vv \kern-0.2em}\left\langle  r\alpha  \right|\otimes \left| {{\Gamma }_{+}} \right\rangle _{\kern-0.1em 66 \kern-0.2em}\left\langle  {{\Gamma }_{-}} \right|\otimes \left| {{\Gamma }_{-}} \right\rangle _{\kern-0.1em 77 \kern-0.2em}\left\langle  {{\Gamma }_{+}} \right| \nonumber\\
		&& +a{{b}^{*}}{{e}^{-\frac{\left( \gamma +4{{\gamma }_{\phi }} \right)\tau }{2}}}\left| \uparrow  \right\rangle _{\kern-0.1em 11 \kern-0.2em}\left\langle  \downarrow  \right|\otimes \left| r\alpha  \right\rangle _{\kern-0.1em vv \kern-0.2em}\left\langle  -r\alpha  \right|\otimes \left| {{\Gamma }_{+}} \right\rangle _{\kern-0.1em 66 \kern-0.2em}\left\langle  -{{\Gamma }_{+}} \right|\otimes \left| {{\Gamma }_{-}} \right\rangle _{\kern-0.1em 77 \kern-0.2em}\left\langle  -{{\Gamma }_{-}} \right| \nonumber\\
		&& +a{{b}^{*}}{{e}^{-\frac{\left( \gamma +4{{\gamma }_{\phi }} \right)\tau }{2}}}\left| \downarrow  \right\rangle _{\kern-0.1em 11 \kern-0.2em}\left\langle  \uparrow  \right|\otimes \left| -r\alpha  \right\rangle _{\kern-0.1em vv \kern-0.2em}\left\langle  r\alpha  \right|\otimes\left| -{{\Gamma }_{-}} \right\rangle _{\kern-0.1em 66 \kern-0.2em}\left\langle  {{\Gamma }_{-}} \right|\otimes \left| -{{\Gamma }_{+}} \right\rangle _{\kern-0.1em 77 \kern-0.2em}\left\langle  {{\Gamma }_{+}} \right| \nonumber\\
		&& +a{{b}^{*}}{{e}^{-\gamma \tau }}\left| \downarrow  \right\rangle _{\kern-0.1em 11 \kern-0.2em}\left\langle  \downarrow  \right|\otimes \left| -r\alpha  \right\rangle _{\kern-0.1em vv \kern-0.2em}\left\langle  -r\alpha  \right|\otimes \left| -{{\Gamma }_{-}} \right\rangle _{\kern-0.1em 66 \kern-0.2em}\left\langle  -{{\Gamma }_{+}} \right|\otimes \left| -{{\Gamma }_{+}} \right\rangle _{\kern-0.1em 77 \kern-0.2em}\left\langle  -{{\Gamma }_{-}} \right| \nonumber\\
		&& +a{{b}^{*}}\left( 1-{{e}^{-\gamma \tau }} \right)\left| \uparrow  \right\rangle _{\kern-0.1em 11 \kern-0.2em}\left\langle  \uparrow  \right|\otimes \left| -r\alpha  \right\rangle _{\kern-0.1em vv \kern-0.2em}\left\langle  -r\alpha  \right|\otimes \left| -{{\Gamma }_{-}} \right\rangle _{\kern-0.1em 66 \kern-0.2em}\left\langle  -{{\Gamma }_{+}} \right|\otimes \left| -{{\Gamma }_{+}} \right\rangle _{\kern-0.1em 77 \kern-0.2em}\left\langle  -{{\Gamma }_{-}} \right| \nonumber\\
		&& +\frac{1}{2}{{\left| b \right|}^{2}}\left| \uparrow  \right\rangle _{\kern-0.1em 11 \kern-0.2em}\left\langle  \uparrow  \right|\otimes \left| r\alpha  \right\rangle _{\kern-0.1em vv \kern-0.2em}\left\langle  r\alpha  \right|\otimes \left| {{\Gamma }_{-}} \right\rangle _{\kern-0.1em 66 \kern-0.2em}\left\langle  {{\Gamma }_{-}} \right|\otimes \left| {{\Gamma }_{+}} \right\rangle _{\kern-0.1em 77 \kern-0.2em}\left\langle  {{\Gamma }_{+}} \right| \nonumber\\
		&& +{{\left| b \right|}^{2}}{{e}^{-\frac{\left( \gamma +4{{\gamma }_{\phi }} \right)\tau }{2}}}\left| \uparrow  \right\rangle _{\kern-0.1em 11 \kern-0.2em}\left\langle  \downarrow  \right|\otimes \left| r\alpha  \right\rangle _{\kern-0.1em vv \kern-0.2em}\left\langle  -r\alpha  \right|\otimes \left| {{\Gamma }_{-}} \right\rangle _{\kern-0.1em 66 \kern-0.2em}\left\langle  -{{\Gamma }_{+}} \right|\otimes \left| {{\Gamma }_{+}} \right\rangle _{\kern-0.1em 77 \kern-0.2em}\left\langle  -{{\Gamma }_{-}} \right| \nonumber\\
		&& +\frac{1}{2}{{\left| b \right|}^{2}}{{e}^{-\gamma \tau }}\left| \downarrow  \right\rangle _{\kern-0.1em 11 \kern-0.2em}\left\langle  \downarrow  \right|\otimes \left| -r\alpha  \right\rangle _{\kern-0.1em vv \kern-0.2em}\left\langle  -r\alpha  \right|\otimes \left| -{{\Gamma }_{+}} \right\rangle _{\kern-0.1em 66 \kern-0.2em}\left\langle  -{{\Gamma }_{+}} \right|\otimes \left| -{{\Gamma }_{-}} \right\rangle _{\kern-0.1em 77 \kern-0.2em}\left\langle  -{{\Gamma }_{-}} \right| \nonumber\\
		&& +\frac{1}{2}{{\left| b \right|}^{2}}\left( 1-{{e}^{-\gamma \tau }} \right) \left| \uparrow  \right\rangle _{\kern-0.1em 11 \kern-0.2em}\left\langle  \uparrow  \right|\otimes \left| -r\alpha  \right\rangle _{\kern-0.1em vv \kern-0.2em}\left\langle  -r\alpha  \right|\otimes \left| -{{\Gamma }_{+}} \right\rangle _{\kern-0.1em 66 \kern-0.2em}\left\langle  -{{\Gamma }_{+}} \right|\otimes \left| -{{\Gamma }_{-}} \right\rangle _{\kern-0.1em 77 \kern-0.2em}\left\langle  -{{\Gamma }_{-}} \right|+\rm{H.c.},
	\end{eqnarray}
where we have defined $\Gamma_\pm=\ket{(t\alpha\pm\beta)/\sqrt{2}}$. Next, the light modes $6$ and $7$ will be reflected by the spin-cavity systems $2$ and $3$, respectively, leading to the state of the system
	\begin{eqnarray}
		{{\hat \rho }_{\rm{out2}}} &=& \hat U_{26}^{CP}\hat U_{37}^{CP}{{\hat \rho }_{{\rm{out1}}}} \otimes \left( {{{\left|  \uparrow  \right\rangle }_2} + {{\left|  \downarrow  \right\rangle }_2}} \right)\left( {{}_2\left\langle  \uparrow  \right| + {}_2\left\langle  \downarrow  \right|} \right) \otimes \left( {{{\left|  \uparrow  \right\rangle }_3} + {{\left|  \downarrow  \right\rangle }_3}} \right)\left( {{}_3\left\langle  \uparrow  \right| + {}_3\left\langle  \downarrow  \right|} \right)\hat U_{37}^{CP\dag }\hat U_{26}^{CP\dag }\nonumber\\
		 &&= \frac{1}{2}{\left| a \right|^2}\left|  \uparrow  \right\rangle _{\kern-0.1em 11 \kern-0.2em}\left\langle  \uparrow  \right| \otimes \left| {r\alpha } \right\rangle _{\kern-0.1em vv \kern-0.2em}\left\langle {r\alpha } \right| \otimes \left( {{{\left| {{\Gamma _ + }} \right\rangle }_8}{{\left|  \uparrow  \right\rangle }_2} + {{\left| { - {\Gamma _ + }} \right\rangle }_8}{{\left|  \downarrow  \right\rangle }_2}} \right)\left( {{}_8\left\langle {{\Gamma _ + }} \right|{}_2\left\langle  \uparrow  \right| + {}_8\left\langle { - {\Gamma _ + }} \right|{}_2\left\langle  \downarrow  \right|} \right)\nonumber\\
		 && \otimes \left( {{{\left| {{\Gamma _ - }} \right\rangle }_9}{{\left|  \uparrow  \right\rangle }_3} + {{\left| { - {\Gamma _ - }} \right\rangle }_9}{{\left|  \downarrow  \right\rangle }_3}} \right)\left( {{}_9\left\langle {{\Gamma _ - }} \right|{}_3\left\langle  \uparrow  \right| + {}_9\left\langle { - {\Gamma _ - }} \right|{}_3\left\langle  \downarrow  \right|} \right)\nonumber\\
		 && + {\left| a \right|^2}{e^{ - \frac{{\left( {\gamma  + 4{\gamma _\phi }} \right)\tau }}{2}}}\left|  \uparrow  \right\rangle _{\kern-0.1em 11 \kern-0.2em}\left\langle  \downarrow  \right| \otimes \left| {r\alpha } \right\rangle _{\kern-0.1em vv \kern-0.2em}\left\langle { - r\alpha } \right| \otimes \left( {{{\left| {{\Gamma _ + }} \right\rangle }_8}{{\left|  \uparrow  \right\rangle }_2} + {{\left| { - {\Gamma _ + }} \right\rangle }_8}{{\left|  \downarrow  \right\rangle }_2}} \right)\left( {{}_8{{\left\langle { - {\Gamma _ - }} \right|}_2}\left\langle  \uparrow  \right| + {}_8\left\langle {{\Gamma _ - }} \right|{}_2\left\langle  \downarrow  \right|} \right)\nonumber\\
		 && \otimes \left( {{{\left| {{\Gamma _ - }} \right\rangle }_9}{{\left|  \uparrow  \right\rangle }_3} + {{\left| { - {\Gamma _ - }} \right\rangle }_9}{{\left|  \downarrow  \right\rangle }_3}} \right)\left( {{}_9\left\langle { - {\Gamma _ + }} \right|{}_3\left\langle  \uparrow  \right| + {}_9\left\langle {{\Gamma _ + }} \right|{}_3\left\langle  \downarrow  \right|} \right)\nonumber\\
		 &&	+ \frac{1}{2}{\left| a \right|^2}{e^{ - \gamma \tau }}\left|  \downarrow  \right\rangle _{\kern-0.1em 11 \kern-0.2em}\left\langle  \downarrow  \right| \otimes \left| { - r\alpha } \right\rangle _{\kern-0.1em vv \kern-0.2em}\left\langle { - r\alpha } \right| \otimes \left( {{{\left| { - {\Gamma _ - }} \right\rangle }_8}{{\left|  \uparrow  \right\rangle }_2} + {{\left| {{\Gamma _ - }} \right\rangle }_8}{{\left|  \downarrow  \right\rangle }_2}} \right)\left( {{}_8\left\langle { - {\Gamma _ - }} \right|{}_2\left\langle  \uparrow  \right| + {}_8\left\langle {{\Gamma _ - }} \right|{}_2\left\langle  \downarrow  \right|} \right)\nonumber\\
		 && \otimes \left( {{{\left| { - {\Gamma _ + }} \right\rangle }_9}{{\left|  \uparrow  \right\rangle }_3} + {{\left| {{\Gamma _ + }} \right\rangle }_9}{{\left|  \downarrow  \right\rangle }_3}} \right)\left( {{}_9\left\langle { - {\Gamma _ + }} \right|{}_3\left\langle  \uparrow  \right| + {}_9\left\langle {{\Gamma _ + }} \right|{}_3\left\langle  \downarrow  \right|} \right)\nonumber\\
		 && + \frac{1}{2}{\left| a \right|^2}\left( {1 - {e^{ - \gamma \tau }}} \right)\left|  \uparrow  \right\rangle _{\kern-0.1em 11 \kern-0.2em}\left\langle  \uparrow  \right| \otimes \left| { - r\alpha } \right\rangle _{\kern-0.1em vv \kern-0.2em}\left\langle { - r\alpha } \right| \otimes \left( {{{\left| { - {\Gamma _ - }} \right\rangle }_8}{{\left|  \uparrow  \right\rangle }_2} + {{\left| {{\Gamma _ - }} \right\rangle }_8}{{\left|  \downarrow  \right\rangle }_2}} \right)\left( {{}_8\left\langle { - {\Gamma _ - }} \right|{}_2\left\langle  \uparrow  \right| + {}_8\left\langle {{\Gamma _ - }} \right|{}_2\left\langle  \downarrow  \right|} \right)\nonumber\\
		 &&	\otimes \left( {{{\left| { - {\Gamma _ + }} \right\rangle }_9}{{\left|  \uparrow  \right\rangle }_3} + {{\left| {{\Gamma _ + }} \right\rangle }_9}{{\left|  \downarrow  \right\rangle }_3}} \right)\left( {{}_9\left\langle { - {\Gamma _ + }} \right|{}_3\left\langle  \uparrow  \right| + {}_9\left\langle {{\Gamma _ + }} \right|{}_3\left\langle  \downarrow  \right|} \right)\nonumber\\
		 &&	+ a{b^ * }\left|  \uparrow  \right\rangle _{\kern-0.1em 11 \kern-0.2em}\left\langle  \uparrow  \right| \otimes \left| {r\alpha } \right\rangle _{\kern-0.1em vv \kern-0.2em}\left\langle {r\alpha } \right| \otimes \left( {{{\left| {{\Gamma _ + }} \right\rangle }_8}{{\left|  \uparrow  \right\rangle }_2} + {{\left| { - {\Gamma _ + }} \right\rangle }_8}{{\left|  \downarrow  \right\rangle }_2}} \right)\left( {{}_8\left\langle {{\Gamma _ - }} \right|{}_2\left\langle  \uparrow  \right| + {}_8\left\langle { - {\Gamma _ - }} \right|{}_2\left\langle  \downarrow  \right|} \right)\nonumber\\
		 &&	\otimes \left( {{{\left| {{\Gamma _ - }} \right\rangle }_9}{{\left|  \uparrow  \right\rangle }_3} + {{\left| { - {\Gamma _ - }} \right\rangle }_9}{{\left|  \downarrow  \right\rangle }_3}} \right)\left( {{}_9\left\langle {{\Gamma _ + }} \right|{}_3\left\langle  \uparrow  \right| + {}_9\left\langle { - {\Gamma _ + }} \right|{}_3\left\langle  \downarrow  \right|} \right)\nonumber\\
		 &&	+ a{b^ * }{e^{ - \frac{{\left( {\gamma  + 4{\gamma _\phi }} \right)\tau }}{2}}}\left|  \uparrow  \right\rangle _{\kern-0.1em 11 \kern-0.2em}\left\langle  \downarrow  \right| \otimes \left| {r\alpha } \right\rangle _{\kern-0.1em vv \kern-0.2em}\left\langle { - r\alpha } \right| \otimes \left( {{{\left| {{\Gamma _ + }} \right\rangle }_8}{{\left|  \uparrow  \right\rangle }_2} + {{\left| { - {\Gamma _ + }} \right\rangle }_8}{{\left|  \downarrow  \right\rangle }_2}} \right)\left( {{}_8\left\langle { - {\Gamma _ + }} \right|{}_2\left\langle  \uparrow  \right| + {}_8\left\langle {{\Gamma _ + }} \right|{}_2\left\langle  \downarrow  \right|} \right)\nonumber\\
		 &&	\otimes \left( {{{\left| {{\Gamma _ - }} \right\rangle }_9}{{\left|  \uparrow  \right\rangle }_3} + {{\left| { - {\Gamma _ - }} \right\rangle }_9}{{\left|  \downarrow  \right\rangle }_3}} \right)\left( {{}_9\left\langle { - {\Gamma _ - }} \right|{}_3\left\langle  \uparrow  \right| + {}_9\left\langle {{\Gamma _ - }} \right|{}_3\left\langle  \downarrow  \right|} \right)\nonumber\\
		 &&	+ a{b^ * }{e^{ - \frac{{\left( {\gamma  + 4{\gamma _\phi }} \right)\tau }}{2}}}\left|  \downarrow  \right\rangle _{\kern-0.1em 11 \kern-0.2em}\left\langle  \uparrow  \right| \otimes \left| { - r\alpha } \right\rangle _{\kern-0.1em vv \kern-0.2em}\left\langle {r\alpha } \right| \otimes \left( {{{\left| { - {\Gamma _ - }} \right\rangle }_8}{{\left|  \uparrow  \right\rangle }_2} + {{\left| {{\Gamma _ - }} \right\rangle }_8}{{\left|  \downarrow  \right\rangle }_2}} \right)\left( {{}_8\left\langle {{\Gamma _ - }} \right|{}_2\left\langle  \uparrow  \right| + {}_8\left\langle { - {\Gamma _ - }} \right|{}_2\left\langle  \downarrow  \right|} \right)\nonumber\\
		 &&	\otimes \left( {{{\left| { - {\Gamma _ + }} \right\rangle }_9}{{\left|  \uparrow  \right\rangle }_3} + {{\left| {{\Gamma _ + }} \right\rangle }_9}{{\left|  \downarrow  \right\rangle }_3}} \right)\left( {{}_9\left\langle {{\Gamma _ + }} \right|{}_3\left\langle  \uparrow  \right| + {}_9\left\langle { - {\Gamma _ + }} \right|{}_3\left\langle  \downarrow  \right|} \right)\nonumber\\
		 &&	+ a{b^ * }{e^{ - \gamma \tau }}\left|  \downarrow  \right\rangle _{\kern-0.1em 11 \kern-0.2em}\left\langle  \downarrow  \right| \otimes \left| { - r\alpha } \right\rangle _{\kern-0.1em vv \kern-0.2em}\left\langle { - r\alpha } \right| \otimes \left( {{{\left| { - {\Gamma _ - }} \right\rangle }_8}{{\left|  \uparrow  \right\rangle }_2} + {{\left| {{\Gamma _ - }} \right\rangle }_8}{{\left|  \downarrow  \right\rangle }_2}} \right)\left( {{}_8\left\langle { - {\Gamma _ + }} \right|{}_2\left\langle  \uparrow  \right| + {}_8\left\langle {{\Gamma _ + }} \right|{}_2\left\langle  \downarrow  \right|} \right)\nonumber\\
		 &&	\otimes \left( {{{\left| { - {\Gamma _ + }} \right\rangle }_9}{{\left|  \uparrow  \right\rangle }_3} + {{\left| {{\Gamma _ + }} \right\rangle }_9}{{\left|  \downarrow  \right\rangle }_3}} \right)\left( {{}_9\left\langle { - {\Gamma _ - }} \right|{}_3\left\langle  \uparrow  \right| + {}_9\left\langle {{\Gamma _ - }} \right|{}_3\left\langle  \downarrow  \right|} \right)\nonumber\\
		 &&	+ a{b^ * }\left( {1 - {e^{ - \gamma \tau }}} \right)\left|  \uparrow  \right\rangle _{\kern-0.1em 11 \kern-0.2em}\left\langle  \uparrow  \right| \otimes \left| { - r\alpha } \right\rangle _{\kern-0.1em vv \kern-0.2em}\left\langle { - r\alpha } \right| \otimes \left( {{{\left| { - {\Gamma _ - }} \right\rangle }_8}{{\left|  \uparrow  \right\rangle }_2} + {{\left| {{\Gamma _ - }} \right\rangle }_8}{{\left|  \downarrow  \right\rangle }_2}} \right)\left( {{}_8\left\langle { - {\Gamma _ + }} \right|{}_2\left\langle  \uparrow  \right| + {}_8\left\langle {{\Gamma _ + }} \right|{}_2\left\langle  \downarrow  \right|} \right)\nonumber\\
		 &&	\otimes \left( {{{\left| { - {\Gamma _ + }} \right\rangle }_9}{{\left|  \uparrow  \right\rangle }_3} + {{\left| {{\Gamma _ + }} \right\rangle }_9}{{\left|  \downarrow  \right\rangle }_3}} \right)\left( {{}_9\left\langle { - {\Gamma _ - }} \right|{}_3\left\langle  \uparrow  \right| + {}_9\left\langle {{\Gamma _ - }} \right|{}_3\left\langle  \downarrow  \right|} \right)\nonumber\\
		 &&	+ \frac{1}{2}{\left| b \right|^2}\left|  \uparrow  \right\rangle _{\kern-0.1em 11 \kern-0.2em} \left\langle  \uparrow  \right| \otimes \left| {r\alpha } \right\rangle_ {\kern-0.1em vv \kern-0.2em}\left\langle {r\alpha } \right| \otimes \left( {{{\left| {{\Gamma _ - }} \right\rangle }_8}{{\left|  \uparrow  \right\rangle }_2} + {{\left| { - {\Gamma _ - }} \right\rangle }_8}{{\left|  \downarrow  \right\rangle }_2}} \right)\left( {{}_8\left\langle {{\Gamma _ - }} \right|{}_2\left\langle  \uparrow  \right| + {}_8\left\langle { - {\Gamma _ - }} \right|{}_2\left\langle  \downarrow  \right|} \right)\nonumber\\
		 &&	\otimes \left( {{{\left| {{\Gamma _ + }} \right\rangle }_9}{{\left|  \uparrow  \right\rangle }_3} + {{\left| { - {\Gamma _ + }} \right\rangle }_9}{{\left|  \downarrow  \right\rangle }_3}} \right)\left( {{}_9\left\langle {{\Gamma _ + }} \right|{}_3\left\langle  \uparrow  \right| + {}_9\left\langle { - {\Gamma _ + }} \right|{}_3\left\langle  \downarrow  \right|} \right)\nonumber\\
		 &&	+ {\left| b \right|^2}{e^{ - \frac{{\left( {\gamma  + 4{\gamma _\phi }} \right)\tau }}{2}}}\left|  \uparrow  \right\rangle _{\kern-0.1em 11 \kern-0.2em}\left\langle  \downarrow  \right| \otimes \left| {r\alpha } \right\rangle _{\kern-0.1em vv \kern-0.2em}\left\langle { - r\alpha } \right| \otimes \left( {{{\left| {{\Gamma _ - }} \right\rangle }_8}{{\left|  \uparrow  \right\rangle }_2} + {{\left| { - {\Gamma _ - }} \right\rangle }_8}{{\left|  \downarrow  \right\rangle }_2}} \right)\left( {{}_8\left\langle { - {\Gamma _ + }} \right|{}_2\left\langle  \uparrow  \right| + {}_8\left\langle {{\Gamma _ + }} \right|{}_2\left\langle  \downarrow  \right|} \right)\nonumber\\
		 &&	\otimes \left( {{{\left| {{\Gamma _ + }} \right\rangle }_9}{{\left|  \uparrow  \right\rangle }_3} + {{\left| { - {\Gamma _ + }} \right\rangle }_9}{{\left|  \downarrow  \right\rangle }_3}} \right)\left( {{}_9\left\langle { - {\Gamma _ - }} \right|{}_3\left\langle  \uparrow  \right| + {}_9\left\langle {{\Gamma _ - }} \right|{}_3\left\langle  \downarrow  \right|} \right)\nonumber\\
		 &&	+ \frac{1}{2}{\left| b \right|^2}{e^{ - \gamma \tau }}\left|  \downarrow  \right\rangle _{\kern-0.1em 11 \kern-0.2em}\left\langle  \downarrow  \right| \otimes \left| { - r\alpha } \right\rangle _{\kern-0.1em vv \kern-0.2em}\left\langle { - r\alpha } \right| \otimes \left( {{{\left| { - {\Gamma _ + }} \right\rangle }_8}{{\left|  \uparrow  \right\rangle }_2} + {{\left| {{\Gamma _ + }} \right\rangle }_8}{{\left|  \downarrow  \right\rangle }_2}} \right)\left( {{}_8\left\langle { - {\Gamma _ + }} \right|{}_2\left\langle  \uparrow  \right| + {}_8\left\langle {{\Gamma _ + }} \right|{}_2\left\langle  \downarrow  \right|} \right)\nonumber\\
		 &&	\otimes \left( {{{\left| { - {\Gamma _ - }} \right\rangle }_9}{{\left|  \uparrow  \right\rangle }_3} + {{\left| {{\Gamma _ - }} \right\rangle }_9}{{\left|  \downarrow  \right\rangle }_3}} \right)\left( {{}_9\left\langle { - {\Gamma _ - }} \right|{}_3\left\langle  \uparrow  \right| + {}_9\left\langle {{\Gamma _ - }} \right|{}_3\left\langle  \downarrow  \right|} \right)\nonumber\\
		 &&	+ \frac{1}{2}{\left| b \right|^2}\left( {1 - {e^{ - \gamma \tau }}} \right)\left|  \uparrow  \right\rangle _{\kern-0.1em 11 \kern-0.2em}\left\langle  \uparrow  \right| \otimes \left| { - r\alpha } \right\rangle _{\kern-0.1em vv \kern-0.2em}\left\langle { - r\alpha } \right| \otimes \left( {{{\left| { - {\Gamma _ + }} \right\rangle }_8}{{\left|  \uparrow  \right\rangle }_2} + {{\left| {{\Gamma _ + }} \right\rangle }_8}{{\left|  \downarrow  \right\rangle }_2}} \right)\left( {{}_8\left\langle { - {\Gamma _ + }} \right|{}_2\left\langle  \uparrow  \right| + {}_8\left\langle {{\Gamma _ + }} \right|{}_2\left\langle  \downarrow  \right|} \right)\nonumber\\
		 &&	\otimes \left( {{{\left| { - {\Gamma _ - }} \right\rangle }_9}{{\left|  \uparrow  \right\rangle }_3} + {{\left| {{\Gamma _ - }} \right\rangle }_9}{{\left|  \downarrow  \right\rangle }_3}} \right)\left( {{}_9\left\langle { - {\Gamma _ - }} \right|{}_3\left\langle  \uparrow  \right| + {}_9\left\langle {{\Gamma _ - }} \right|{}_3\left\langle  \downarrow  \right|} \right)+\rm{H.c.}.
		\end{eqnarray}		
Then, after applying a $\pi/2$ transformation to the auxiliary qubits, both the auxiliary NV spins and the light modes will be detected. If the measurement results are $
\left| \uparrow \right. \rangle _{\kern -0.1em 22\kern -0.2em}\left. \langle \uparrow \right|\text{,}\left| \uparrow \right. \rangle _{\kern -0.1em 33\kern -0.2em}\left. \langle \uparrow \right|,\left| \Gamma _+ \right. \rangle _{\kern -0.1em 88\kern -0.2em}\left. \langle \Gamma _+ \right|\text{,}$ and $\left| \Gamma _- \right. \rangle _{\kern -0.1em 99\kern -0.2em}\left. \langle \Gamma _- \right|
$
, then the state $\hat \rho_{\rm{out2}}$ of the system will collapse to
\begin{eqnarray}
\hat{\rho}_{\mathrm{out}3}&=&\mathcal{N}\left[ \left| a \right|^2\left| \uparrow \right. \rangle _{\kern -0.1em 11\kern -0.2em}\left. \langle \uparrow \right|\otimes \left| r\alpha \right. \rangle _{\kern -0.1em vv\kern -0.2em}\left. \langle r\alpha \right|+ab^*e^{-\frac{\left( \gamma +4\gamma _{\phi} \right) \tau}{2}}\left| \uparrow \right. \rangle _{\kern -0.1em 11\kern -0.2em}\left. \langle \downarrow \right|\otimes \left| r\alpha \right. \rangle _{\kern -0.1em vv\kern -0.2em}\left. \langle -r\alpha \right| \right.
\nonumber\\
&&+ba^*e^{-\frac{\left( \gamma +4\gamma _{\phi} \right) \tau}{2}}\left| \downarrow \right. \rangle _{\kern -0.1em 11\kern -0.2em}\left. \langle \uparrow \right|\otimes \left| -r\alpha \right. \rangle _{\kern -0.1em vv\kern -0.2em}\left. \langle r\alpha \right|+\left| b \right|^2e^{-\gamma \tau}\left| \downarrow \right. \rangle _{\kern -0.1em 11\kern -0.2em}\left. \langle \downarrow \right|\otimes \left| -r\alpha \right. \rangle _{\kern -0.1em vv\kern -0.2em}\left. \langle -r\alpha \right|
\nonumber\\
&&\left. +\left| b \right|^2\left( 1-e^{-\gamma \tau} \right) \left| \uparrow \right. \rangle _{\kern -0.1em 11\kern -0.2em}\left. \langle \uparrow \right|\otimes \left| -r\alpha \right. \rangle _{\kern -0.1em vv\kern -0.2em}\left. \langle -r\alpha \right| \right],
\end{eqnarray}	
where $\mathcal{N}$ is the normalization. Note that the final spin state is entangled with the vacuum mode that should be traced out. To do so, we express the vacuum mode in the position basis $\left| {\pm r\alpha } \right\rangle _{\kern-0.1em vv \kern-0.2em}\left\langle {\pm r\alpha } \right| = {\pi ^{ - {1 \mathord{\left/
					{\vphantom {1 2}} \right.
					\kern-\nulldelimiterspace} 2}}}\int {dx{e^{ - \frac{{{{(x \mp r\alpha )}^2}}}{2}}}} \int {dx'{e^{ - \frac{{{{(x' \mp r\alpha )}^2}}}{2}}}}\left| x \right\rangle \left\langle {x'} \right|$, then the trace over the vacuum mode results in
\begin{eqnarray}
\hat{\rho}_f&=&\text{Tr}_v\left( \hat{\rho}_{\mathrm{out}3} \right)
=\int{dy}\langle y|\hat{\rho}_{\mathrm{out}3}|y\rangle
\nonumber\\
&=&\left[ \left| a \right|^2+\left| b \right|^2\left( 1-e^{-\gamma \tau} \right) \right] \left| \uparrow \right. \rangle _{\kern -0.1em 11\kern -0.2em}\left. \langle \uparrow \right|+ab^*e^{-\frac{\left( 4\gamma _{\phi}+\gamma \right) \tau}{2}-r^2\alpha ^2}\left| \uparrow \right. \rangle _{\kern -0.1em 11\kern -0.2em}\left. \langle \downarrow \right|
\nonumber\\
&&+ba^*e^{-\frac{\left( 4\gamma _{\phi}+\gamma \right) \tau}{2}-r^2\alpha ^2}\left| \downarrow \right. \rangle _{\kern -0.1em 11\kern -0.2em}\left. \langle \uparrow \right|+\left| b \right|^2e^{-\gamma \tau}\left| \downarrow \right. \rangle _{\kern -0.1em 11\kern -0.2em}\left. \langle \downarrow \right|.\label{eqa4}
\end{eqnarray}
In deriving this equation, we have used the results
\begin{eqnarray}
\int { dy\left\langle {y}
		\mathrel{\left | {\vphantom {y {r\alpha }}}
			\right. \kern-\nulldelimiterspace}
		{{r\alpha }} \right\rangle \left\langle {{r\alpha }}
		\mathrel{\left | {\vphantom {{r\alpha } y}}
			\right. \kern-\nulldelimiterspace}
		{y} \right\rangle } &=& \int {dy\left\langle {y}
		\mathrel{\left | {\vphantom {y { - r\alpha }}}
			\right. \kern-\nulldelimiterspace}
		{{ - r\alpha }} \right\rangle \left\langle {{ - r\alpha }}
		\mathrel{\left | {\vphantom {{ - r\alpha } y}}
			\right. \kern-\nulldelimiterspace}
		{y} \right\rangle } = {\pi ^{ - {1 \mathord{\left/
					{\vphantom {1 2}} \right.
					\kern-\nulldelimiterspace} 2}}}\int {dy{e^{ - {{(y - r\alpha )}^2}}}}  = 1,\nonumber\\
\int {dy\left\langle {y}
		\mathrel{\left | {\vphantom {y {r\alpha }}}
			\right. \kern-\nulldelimiterspace}
		{{r\alpha }} \right\rangle \left\langle {{ - r\alpha }}
		\mathrel{\left | {\vphantom {{ - r\alpha } y}}
			\right. \kern-\nulldelimiterspace}
		{y} \right\rangle } &=& \int {dy\left\langle {y}
		\mathrel{\left | {\vphantom {y { - r\alpha }}}
			\right. \kern-\nulldelimiterspace}
		{{ - r\alpha }} \right\rangle \left\langle {{r\alpha }}
		\mathrel{\left | {\vphantom {{r\alpha } y}}
			\right. \kern-\nulldelimiterspace}
		{y} \right\rangle } = {\pi ^{ - {1 \mathord{\left/
					{\vphantom {1 2}} \right.
					\kern-\nulldelimiterspace} 2}}}\int {dy{e^{ - {y^2}}}} {e^{ - {{(r\alpha )}^2}}} = {e^{ - {{(r\alpha )}^2}}}.\nonumber
\end{eqnarray}	

Eq. (\ref{eqa4}) is the result given in Eq. (\ref{eq23}) in the main text.
\end{widetext}

	\bibliography{ref}
	
\end{document}